\documentclass[useAMS,usenatbib]{mn2e}
\usepackage{amsmath}
\usepackage{amssymb}
\usepackage{multirow}
\usepackage{graphicx}
\usepackage{epstopdf}

\usepackage{hyperref}
 \hypersetup{
  colorlinks=true,
    linkcolor=blue,
 citecolor=blue,      
  urlcolor=magenta,
}

\usepackage{color}

\def\simlt{\lower.5ex\hbox{$\; \buildrel < \over \sim \;$}}
\def\simgt{\lower.5ex\hbox{$\; \buildrel > \over \sim \;$}}

\newcommand{\bd}{\begin{displaymath}}
\newcommand{\ed}{\end{displaymath}}
\newcommand{\be}{\begin{equation}}
\newcommand{\ee}{\end{equation}}
\newcommand{\beqa}{\begin{eqnarray}}
\newcommand{\eeqa}{\end{eqnarray}}

\title[21-cm Global Signal Emulator] {Emulating the Global 21-cm
  Signal from Cosmic Dawn and Reionization}  \author[Cohen et al.] {Aviad Cohen$^{1}$, Anastasia Fialkov$^{2}$\thanks{E-mail: afialkov@ast.cam.ac.uk}, Rennan
        Barkana$^{1}$, Raul A. Monsalve$^{3,4,5}$ \\ $^{1}$ Raymond and
        Beverly Sackler School of Physics and Astronomy, Tel Aviv
        University, Tel Aviv 69978, Israel\\ 
        $^{2}$ Institute of Astronomy, University of Cambridge, Madingley Road, Cambridge CB3 0HA, UK\\
        $^{3}$ Department of Physics and McGill Space Institute, 3600 Rue University, QC H3A 2T8, Canada\\
        $^{4}$ School of Earth and Space Exploration, Arizona State University, Tempe, AZ 85287, USA
 \\ $^{5}$ Facultad de Ingenier\'ia, Universidad Cat\'olica de la Sant\'isima Concepci\'on, Alonso de Ribera 2850, Concepci\'on, Chile
  }%%%

\begin{document}
\pagerange{\pageref{firstpage}--\pageref{lastpage}} \pubyear{2018}
\maketitle

\label{firstpage}

\begin{abstract} 

The 21-cm signal of neutral hydrogen is a sensitive probe of the Epoch
of Reionization (EoR), Cosmic Dawn and the Dark Ages. Currently operating
radio telescopes have ushered in a data-driven era of 21-cm cosmology,
providing the first constraints on the astrophysical properties of
sources that drive this signal. However, extracting astrophysical
information from the data is highly non-trivial and requires the rapid
generation of theoretical templates over a wide range of astrophysical
parameters. To this end emulators are often employed, with previous
efforts focused on predicting the power spectrum. In this work we
introduce {\sc 21cmGEM} -- the first emulator of the global 21-cm signal from Cosmic
Dawn and the EoR. The smoothness of the output
signal is guaranteed by design. We train neural networks to predict
the cosmological signal using a database of $\sim$30,000 simulated signals which were created by varying  seven astrophysical parameters:   the star
formation efficiency and the minimal mass of star-forming halos; the efficiency of the first X-ray sources and their spectrum parameterized by spectral index and the low energy cutoff; the mean free path of ionizing photons and the CMB  optical depth.
  We test the
performance with a set of $\sim$2,000 simulated signals, showing that
the relative error in the prediction has an r.m.s.\ of 0.0159.  The
algorithm is efficient, with a running time per parameter set of 0.16
sec.  Finally, we use the
database of models to check the robustness of relations between the features of
the global signal and the astrophysical parameters that we previously
reported. 
\end{abstract}


\section{Introduction}
\label{Sec:Intro}

The exploration of the Universe out to times earlier than the point of
complete reionization is rapidly advancing. One of the most
informative probes of these epochs is the 21-cm line produced by
hydrogen atoms in the neutral intergalactic medium (IGM) at redshifts
$z>6$. This line redshifts to frequencies below 200 MHz and can be
detected by low-frequency radio telescopes. Global 21-cm experiments
measure the spectrum of this line averaged over the sky. The first
tentative detection of the Cosmic Dawn signal was recently made by the
Low-Band implementation of the Experiment to Detect the Global EoR Signature
\citep[EDGES,][]{Bowman:2018}. Other global 21-cm experiments,
including the Large-Aperture Experiment to Detect the Dark Ages
\citep[LEDA,][]{Bernardi:2016, Price:2018}, the EDGES High-Band 
\citep{Bowman:2010,Monsalve:2017, Monsalve:2018, Monsalve:2019}, and the Shaped Antenna measurement
of the background RAdio Spectrum
\citep[SARAS,][]{Singh:2017,Singh:2017b}, provide upper limits on the
signal from Cosmic Dawn and the Epoch of Reionization (EoR), ruling
out some extreme astrophysical scenarios.  A parallel effort is being
made by interferometric radio arrays that are placing upper limits on
the fluctuations of the 21-cm signal, including the Donald C.\ Backer
Precision Array for Probing the Epoch of Reionization 
\citep[PAPER,][]{Kolopanis:2019}, the Low Frequency Array
\citep[LOFAR,][]{Patil:2017, Gehlot:2019, Mertens:2020}, the Giant Metrewave Radio Telescope
\citep[GMRT,][]{Paciga:2013}, the Murchison Widefield Array
\citep[MWA,][]{Beardsley:2016, Barry:2019, Li:2019, Trott:2020}, and the Owens Valley Radio Observatory Long Wavelength Array \citep[OVRO-LWA,][]{Eastwood:2019}. The most recent upper limit reported by LOFAR \citep{Mertens:2020} made it possible to place (weak) upper limits on the temperature of the neutral gas and ionization state of the Universe at $z=9.1$ \citep{Ghara:2020, Mondal:2020}.   
Upcoming arrays, including the
Hydrogen Epoch of Reionization Array \citep[HERA,][]{DeBoer:2017}, the
Square Kilometer Array \citep[SKA,][]{Koopmans:2015} and the New
Extension in Nancay Upgrading LOFAR \citep[NenuFAR,][]{Zarka:2012},
will provide measurements of the power spectrum over a wide range of
scales and redshifts.

The 21-cm signal is driven by both astrophysical and cosmological
processes and is thus a unique probe of the early Universe. The
amplitude of the 21-cm line observed against the radio background
radiation, normally assumed to be the Cosmic Microwave Background
\citep[CMB; however, see][]{Feng:2018, Ewall-Wice:2018, Ewall-Wice:2019}, depends on
the abundance of neutral hydrogen atoms as well as on the contrast
between the spin temperature, $T_S$ (the excitation temperature of the
21-cm transition), and the temperature of the background, $T_{\rm
  rad}$. The former is driven to the kinetic temperature of the gas,
$T_K$, by collisions as well as via absorption and re-emission of
stellar Ly$\alpha$ photons \citep[the Wouthuysen-Field, WF, 
  coupling,][]{Wouthuysen:1952, Field:1958}. In the absence of
collisions and/or Ly$\alpha$ radiation the spin temperature is driven
to the temperature of the background. The gas is seen in absorption
against the background if $T_S<T_{\rm rad}$ (usually during the Dark
Ages and Cosmic Dawn). Once the population of the first X-ray sources
builds up and heats the IGM above the temperature of the background,
the gas is seen in emission. In the course of reionization the
abundance of neutral hydrogen atoms decreases and the IGM signal
gradually vanishes. Overall, the signal measures properties of star
formation, the abundance and luminosity of UV and X-ray sources, and
possibly, properties of dark matter if the latter has an effect on the
thermal and ionization histories of the gas
\citep[e.g.,][]{Barkana:2018, Fialkov:2018, DAmico:2018, Munoz:2015,
  Evoli:2014, Tashiro:2014}.

Our currently limited knowledge about primordial star and black hole
formation translates into large uncertainties in the predicted 21-cm
signal. As a result, a wide space of astrophysical parameters should
be explored when predicting the 21-cm signature. Because full-scale
numerical simulations are prohibitively expensive, alternative
techniques, such as fast algorithms, emulators, or machine learning
methods, are often employed to walk through the allowed space of
astrophysical signals \citep[e.g.,][]{Greig:2015, Greig:2017b, Schmit:2018,
  Shimabukuro:2017}.  The effort has so far focused on the
power spectrum of the 21-cm signal: \citet{Greig:2015} presented
21CMMC -- a Monte-Carlo Markov Chain (MCMC) tool which returns three
reionization parameters (the mean free path of ionizing photons, the
minimum temperature of star forming halos and the ionizing efficiency
of sources) given power spectrum measurements  \citep[similar efforts include works by][]{Liu:2016, Hassan:2017}. Because X-ray heating
might play an important role during the EoR \citep{Mesinger:2013,
  Fialkov:2014b}, 21CMMC was recently extended to include three
heating parameters: the bolometric luminosity of X-ray sources per
unit star formation rate, as well as the low-energy cutoff and the
slope of the X-ray spectral energy distribution
\citep{Greig:2017b}. \citet{Shimabukuro:2017} took a different
approach to find the best fit reionization parameters given power
spectrum measurements: artificial Neural Networks (NNs) were trained
on the data from 70 EoR simulations performed using the 21cmFAST code
\citep{Mesinger:2011}. The performance of the algorithm was tested on
an additional set of 54 simulations.  \citet{Schmit:2018} used NNs to
emulate the power spectra generated by 21cmFAST and found a good
agreement with 21CMMC.  \citet{Jennings:2019}  compared five different machine learning techniques for emulating the power spectrum of models generated with the code SimFast21 \citep{Santos:2010}. Finally, \citet{Kern:2017} presented a more sophisticated emulator based on Gaussian processes, which could be applied to a broad range of problems.
They demonstrate the performance on a six-parameter model for the 21-cm signal including reionization and heating parameters as well as five additional cosmological parameters. With the exception of \citet{Kern:2017}, all the above-mentioned tools are designed to reconstruct the parameters
from a 21-cm power spectrum measurement. Similar tools specifically designed for the global 21-cm signal are lacking.

The recently reported results from EDGES Low-Band 
\citep{Bowman:2018} revealed an anomalously strong and narrow
absorption feature at $\sim 78$ MHz which, if truly of cosmological
origin, cannot be explained by the standard astrophysical model
outlined above.  Even though concerns about the signal being of cosmological origin have been expressed in the literature \citep[it could be a result of an uncompensated systematic error or be imprinted by the Galactic foregrounds,][]{Hills:2018, Sims:2019, Singh:2019, Spinelli:2019}, several exotic theories have been suggested to explain this signal. One possible explanation is that dark matter scattered
off baryons, draining energy from the gas and leading to its over-cooling \citep[e.g.,][]{Barkana:2018,Fialkov:2018, Munoz:2015, Munoz:2018}. Another explanation
invoked in the literature requires the existence of a strong radio
background in addition to the CMB. Such an excess could be created by
an anomalously bright population of high-redshift black holes at
$z\sim 20$ \citep{Bowman:2018, Feng:2018, Ewall-Wice:2018, Ewall-Wice:2019}. As we
await independent observational confirmation of the intriguing EDGES
result, it is important to keep studying both the standard picture and
exotic scenarios. In this paper we explore a wide range of standard
astrophysical scenarios. We use a large dataset of models, which cover
the widest astrophysical parameter space (see the next section), to
develop a 21-cm global emulator ({\sc 21cmGEM}) for the first time. Given a set of
seven astrophysical parameters, the emulator makes a prediction for
the global 21-cm signal over a wide redshift range ($z=5-50$) that
includes both the EoR and Cosmic Dawn. Although our models do not
capture the EDGES absorption feature, the algorithm developed here
could be applied to a revised set of models with additional physics.
{\sc 21cmGEM}, along with the global signals that were used to create the emulator, is  publicly available at \url{https://www.ast.cam.ac.uk/~afialkov/Publications.html}. The tool has recently been employed to derive constraints on astrophysical parameters using the EDGES
High-Band spectrum\footnote{Note that we previously referred to the emulator as  {\sc Global21cm}  \citep{Monsalve:2019}.} \citep[90–190~MHz,][]{Monsalve:2019}.

This paper is organized as follows. In Section~\ref{Sec:Model} we
detail our seven-parameter astrophysical model. In
Section~\ref{Sec:Data} we describe the simulation, the limits on the
astrophysical parameter space, and the database of $\sim 30,000$
models. We also re-examine consistency relations between the
astrophysical parameters and the features of the global signal first
derived by \citet{Cohen:2017}.  The design of the emulator is outlined
in Section~\ref{Sec:Interp}, and its performance assessed. Finally, we
summarize our results in Section~\ref{Sec:Summ}.

\section{The high-redshift Universe}
\label{Sec:Model}

The 21-cm signal from Cosmic Dawn and the EoR is driven by several
astrophysical processes including star formation, heating and
ionization.  To produce the 21-cm signal we use our semi-numerical
method \citep[e.g.,][]{Visbal:2012,Fialkov:2014b} which generates realizations
of the universe in large cosmological volumes (384$^3$ comoving
Mpc$^3$) and over a large redshift range ($z = 5-60$). The simulation
follows the hierarchical growth of structure \citep[including effects of the relative
streaming velocity between dark matter and gas,][]{Tseliakhovich:2010}, tracks star formation
(averaged over a 3~Mpc scale) and follows the evolution of X-ray,
Ly$\alpha$, Lyman-Werner (LW, 11.2-13.6 eV) and ionizing radiative
backgrounds. The simulation takes into account the effect of the
relative streaming velocity on star formation, as well as the effect
of the LW radiation and of the photoheating feedback on star formation
(see details below). We parameterize the high-redshift astrophysics
using seven key parameters: the star formation efficiency ($f_*$), the
minimum circular velocity of star-forming halos ($V_c$), the X-ray
radiation efficiency ($f_X$), power-law slope ($\alpha$) and low
energy cutoff ($\nu_{\rm min}$) of the X-ray spectral energy
distribution (SED), the mean free path of ionizing photons ($R_{\rm
  mfp}$) and the CMB optical depth ($\tau$).

\subsection{Star formation}
The simulation takes into account the effect of radiative and
mechanical feedback processes on star formation. Star formation is
possible in dark matter halos that are massive enough to enable
efficient cooling of the in-falling gas
\citep[e.g.,][]{Tegmark:1997}. We use the threshold mass, or,
equivalently (at a given redshift), the minimum circular velocity of
star forming halos, as one of the free parameters. The lowest
temperature coolant in the early Universe is molecular hydrogen, which
allows stars to form in halos more massive than $M^{\rm
  molecular}_{\rm min} \sim 10^5 M_\odot$, or with circular velocity
larger than $V_c = 4.2$ km s$^{-1}$ \citep[e.g.,][]{Tegmark:1997,
  Barkana:2001, Abel:2002, Bromm:2002, Yoshida:2003}.  LW radiation produced by
the first stars eventually halts star formation in molecular cooling
halos \citep{Haiman:1997}, shifting it to more massive atomic cooling
halos of $M^{\rm atomic}_{\rm min} \sim 10^7 M_\odot$ \citep[$V_c =
  16.5$ km
  s$^{-1}$,][]{Haiman:2000,Machacek:2001,Wise:2007,O'Shea:2008}. The
timing and duration of this transition is affected by uncertainties in
the efficiency of the LW feedback \citep{Visbal:2014,Schauer:2015}. In
addition, star formation in low-mass halos is modulated by the
relative streaming velocity between dark matter and baryons
\citep[e.g.,][]{Tseliakhovich:2010,Dalal:2010,Fialkov:2012, Schauer:2019}. On
the other hand, the minimum cooling mass can rise above the atomic
cooling threshold via feedback mechanisms such as supernova explosions
\citep[e.g.,][]{Wyithe:2013}. At lower redshifts, when reionization
becomes significant and the gas in the IGM is heated above $10^4$ K,
photoheating feedback becomes important. This feedback mechanism
prevents further accretion of gas onto halos below $10^8-10^9 M_\odot$
\citep[$V_c$ up to $\sim 75$ km s$^{-1}$, e.g.,][]{Rees:1986,
  Weinberg:1997, Navarro:2000, Sobacchi:2013, Cohen:2016}.  
In our parameter study we explored the values of $V_c$ between $4.2$ and $100$ km s$^{-1}$. 
However,  the emulator was optimized for the starting value of $V_c$ (before
additional feedbacks are imposed) in the $4.2-76.5$ km s$^{-1}$ range.

Another free parameter in the simulation is the fraction of gas in dark matter halos that is converted into stars,
referred to as the star formation efficiency. In general, this quantity depends on the
halo mass.  At low redshifts observations find a mass dependent star formation efficiency, e.g., \citet{Behroozi:2019} show the evolution of stellar mass in halos above $M\gtrsim 10^{10} M_\odot$ and at $0\lesssim z \lesssim 10$. The observed star formation efficiency peaks at a value of a few percent in halos of $\sim 2.8\times 10^{11}M_\odot$
\citep[e.g.,][]{Mirocha:2017, Behroozi:2019}, as it is regulated by feedback
mechanisms, and the process is less efficient in both higher-mass and
lower-mass halos. Such trends were  recently incorporated  in simulations of reionization with applications to synergies between the 21-cm signal and galaxy surveys with {\it James Webb Space Telescope}  at $z\lesssim 10$ \citep{Mirocha:2017, Park:2020}. However, due to the lack of observations at higher redshifts and lower halo masses, applying such models to our work would require considerable extrapolations. This is because the Cosmic Dawn signal is driven by dark matter halos of $10^5-10^8 M_\odot$ which can start forming stars as early as $z\sim 40$. 

 Star formation in the low-mass halos characteristic
of the high redshift Universe is virtually unconstrained by
observations, while numerical simulations yield a large scatter,
finding an efficiency of a few percent or much lower \citep{Jeon:2014,
  Wise:2014, OShea:2014}. Therefore, to parameterize the process of star
formation in our simulations we assume constant star formation
efficiency in halos heavier than the atomic cooling mass (and this
value we designate $f_*$), while in lower mass halos a logarithmic
cutoff in the efficiency is employed 

\[f_*(M) = \left\{
  \begin{array}{lr}
   f_* & M^{\rm atomic}_{\rm min}<M,\\
   f_*\frac{\log(M/M_{\rm min})}{\log(M^{\rm atomic}_{\rm min}/M_{\rm min})} & M_{\rm min}<M<M^{\rm atomic}_{\rm min},\\
   0 & {\rm otherwise},
  \end{array}
\right.
\] 
where $M_{\rm min}$ corresponds to the cutoff circular velocity $V_c$ \citep[see][for more details]{Cohen:2017}.  We vary $f_*$ between 0.0001 and 0.5. 

 \subsection{Heating}
The least constrained component of the modeling is the set of
properties of the first X-ray sources that heat up the cosmic gas.
The most plausible sources that dominate the X-ray radiative
background at high redshifts are X-ray binaries
\citep[XRBs,][]{Mirabel:2011, Fragos:2013}; however, other candidates
have also been discussed in the literature, including hot gas in
galaxies, mini-quasars \citep{Madau:2004}, X-rays produced via inverse
Compton scattering of the CMB photons off electrons accelerated by
supernovae \citep{Oh:2001}, or more exotic scenarios such as dark
matter annihilation \citep[e.g.,][]{Cirelli:2009, Liu:2019}.

The SED of the early X-ray sources is a
key astrophysical parameter \citep{Fialkov:2014b} and might strongly affects the 21-cm signal from both the EoR and Cosmic Dawn. The effect of hard
X-ray sources with energy around 2~keV on the thermal and ionization
histories (and, thus, on the resulting 21-cm signal) is significantly
different from that of soft sources with energies of $\sim 0.5$~keV:
soft sources generate strong fluctuations on relatively small scales
(up to a few tens of comoving Mpc) in the gas temperature and,
subsequently, in the 21-cm intensity; on the other hand, hard sources
produce a more homogeneous and less efficient heating, generating mild
fluctuations on larger scales ($>100$~comoving Mpc). XRBs, as well as
miniquasars, have a hard spectral energy distribution \citep[see the
  discussion in][]{Fialkov:2016} that peaks at a few keV, while other
sources can have softer SEDs. Absorption of soft X-rays with energy
lower than $\nu_{\rm min}$ (typically of $\sim 0.1-0.5$ keV) by dust
in the host galaxy could contribute to effective hardening of X-ray
SEDs \citep{Fragos:2013}.  We parameterize the X-ray SED by a
power-law of the slope $\alpha$ (i.e., $d\log(E_X)/d\log(\nu) =
-\alpha$) and a low-frequency cutoff $\nu_{\rm min}$. Since there is
significant degeneracy between these two parameters, we vary $\alpha$
only slightly (in the range: $\alpha = 1 - 1.5$), and $\nu_{\rm min}$
in the wide range of $0.1-3$ keV.

In addition to the shape of the SED, the total X-ray luminosity of
sources is the other important parameter. Here we adopt the standard
expression for the luminosity per star formation rate
\citep[$L_{X}-$SFR relation, see ][ for more details]{Fialkov:2014b,
  Cohen:2017} inferred from low-redshift observations of nearby
starburst galaxies and XRBs
\citep{Grimm:2003,Gilfanov:2004,Mineo:2012}:
\begin{equation}
\label{eq:XraySFR}
\frac{L_X}{\rm SFR}=3\times10^{40}f_X  \rm~erg~s^{-1}~M_\odot^{-1}~yr.
\end{equation}
In the above expression $L_X$ is the bolometric luminosity, and $f_X$
is the (constant) X-ray efficiency of sources, which we use as the
third X-ray parameter.  The standard normalization for XRBs (with
$f_X=1$) takes into account an order-of-magnitude increase in the
$L_{X}-$SFR relation in the low-metallicity environments expected at
high redshifts \citep{Fragos:2013}. The high-redshift $f_X$ is poorly
constrained: A model-dependent upper limit of $f_X \sim 10-1000$ can
be derived using the measurement of the unresolved cosmic X-ray
background \citep{Fialkov:2016}; a lower (also model-dependent) limit
of $f_X\sim 0.001$ is hinted at by 21-cm experiments
\citep{Singh:2017, Singh:2017b, Monsalve:2018, Monsalve:2019, Mondal:2020}. 
To explore the parameter space we vary $f_X$ between $0$ and $1000$. 
However,  the emulator was optimized in the range of $f_X$ between 0 and 10.

\subsection{Reionization}
\label{Sec:eor1}

We parameterize the process of reionization with two parameters: The
first parameter is the mean free path of ionizing photons, $R_{\rm
  mfp}$, which we vary between 10 and 50 comoving Mpc
\citep{Alvarez:2012, Greig:2015}. This parameter approximately
quantifies the effect of dense small-scale absorption systems in that
it is the mean free path of ionizing photons in a large-scale ionized
bubble. In practice it is set as an upper limit on the distance to
sources that can participate in the reionization of a given cell. 

The
second EoR parameter is the ionizing efficiency of sources, defined as
 \begin{equation}
 \zeta=f_* f_{\rm esc} N_{\rm ion}\frac{1}{1+\bar{n}_{\rm rec}}\ ,
 \label{Eq:zeta}
\end{equation}   
where $f_{\rm esc}$ is the fraction of ionizing photons that escape
into the IGM, $\bar{n}_{\rm rec}$ is the mean number of recombinations
per ionized hydrogen atom, and $N_{\rm ion}$ is the number of ionizing
photons produced per stellar baryon. Given a star-formation history  (i.e., fixing all the rest of the parameters  
$[f_*,~V_c,~f_X,~\alpha,~\nu_{\rm min},~R_{\rm mfp}]$), and assuming a mass-independent ionizing efficiency, there is a
one-to-one correspondence between $\zeta$ and the CMB optical depth. 
Because
$\tau$ (rather than $\zeta$) is directly probed by the CMB experiments
\citep[specifically by the {\it Planck} satellite,][]{Planck:2016b},
we choose to work with $\tau$ as the free parameter. 
 In our parameter exploration we varied $\tau$ between $\sim 0.04$ and $\sim 0.2$. However, high values of $\tau$ are ruled out \citep[e.g.,][]{Planck:2016b}, and we find it difficult to produce $\tau$ below 0.055  and still be consistent with observational constraints (see below). Therefore, the emulator has been optimized for $\tau$ in the range between 0.055 and 0.1.
 
\begin{figure*}
	\centering
	\includegraphics[width=3.5in]{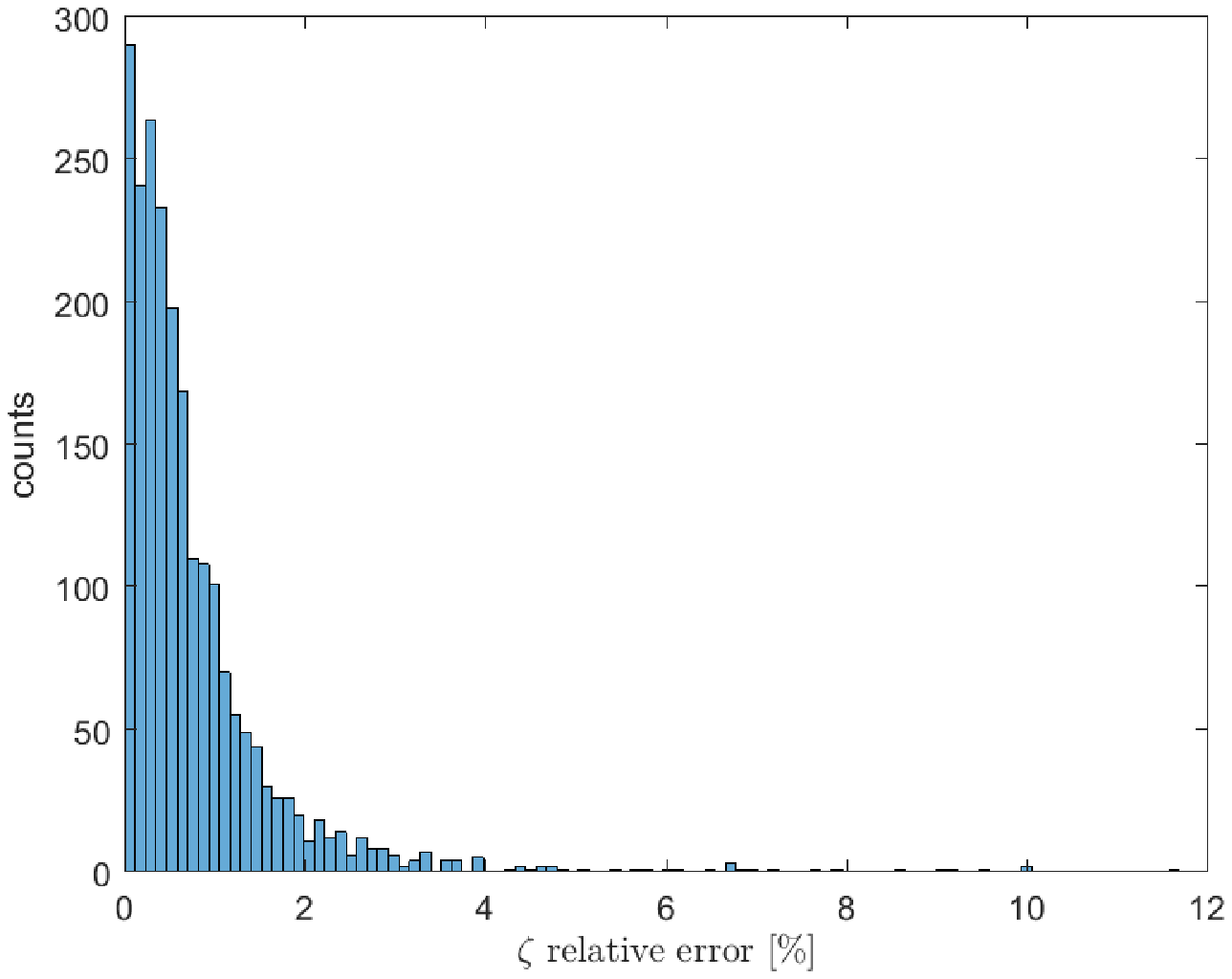}\includegraphics[width=3.5in]{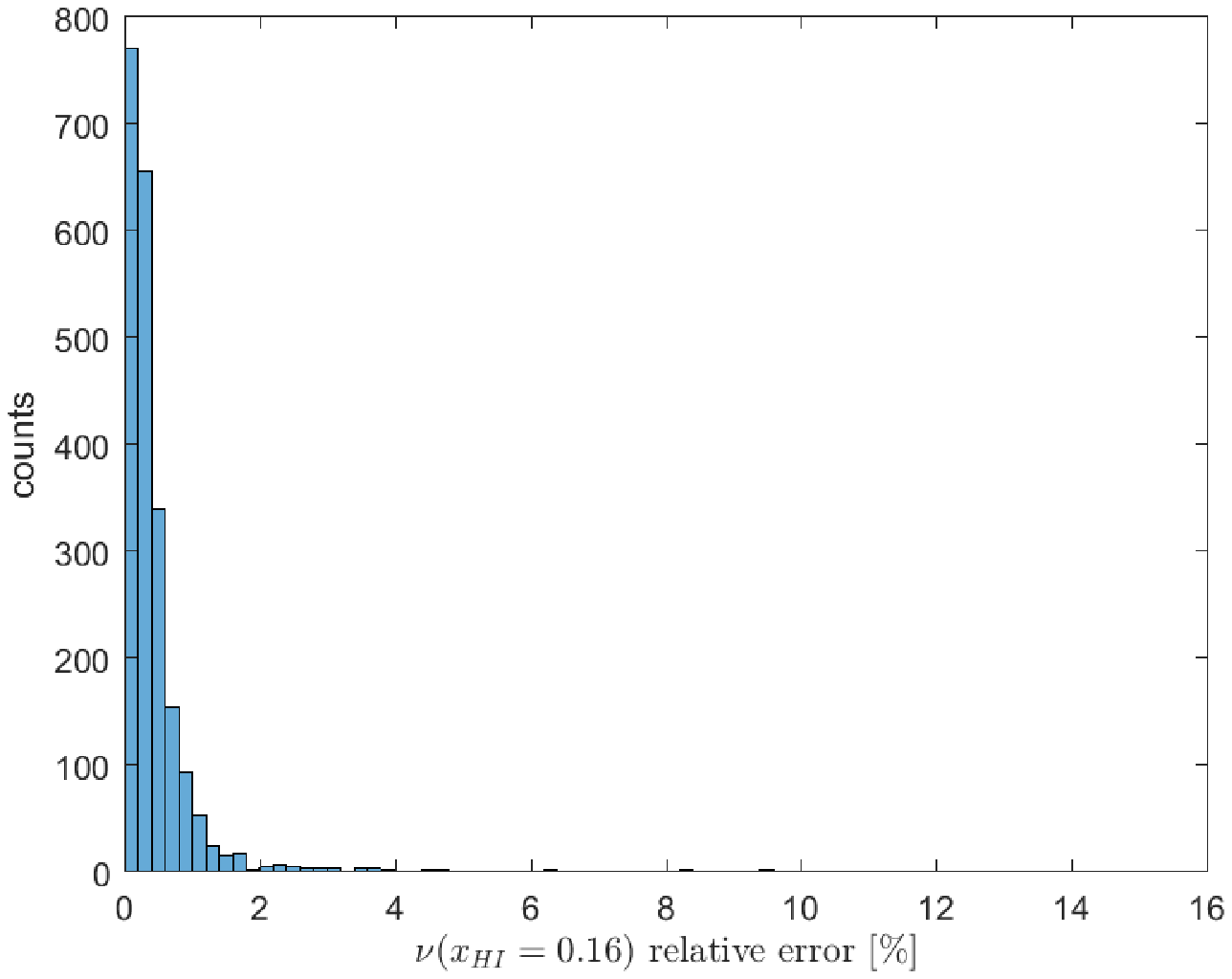}
	\caption{Left: Histogram of the relative error in the prediction of
          $\zeta$ based on  7 model parameters $[f_*,~V_c,~f_X,~\alpha,~\nu_{\rm min},~R_{\rm mfp},\tau]$.
          The total number of test cases was 2,186. We find that 76\% of the
cases have a relative error smaller than 1\%. Right: Histogram of the relative error in the prediction of
          $\nu_{16\%}$, the frequency (which is a measure of redshift)
         at which the neutral fraction reaches 16\%.  92\% of cases have a relative error smaller than 1\%.}
	\label{fig:zeta-hist}
\end{figure*}

The non-linear mapping between $\tau$ and $\zeta$, which is a function
of the other input astrophysical parameters, is carried out using a NN
which was trained on a set of 27,455 cases and tested with 2,186
cases.   This NN has 7 input model parameters $[f_*,~V_c,~f_X,~\alpha,~\nu_{\rm min},~R_{\rm mfp},\tau]$ (and, thus, 7 input neurons),
 one hidden layer of 40 neurons and 1 output, $\zeta$. The Levenberg-Marquardt algorithm \citep{Levenberg, Marquardt} was used 
 to minimize the
mean-square error between the true value provided by the training
dataset and the value predicted by the network.  We evaluate the
performance of the NN by quantifying its accuracy in predicting
$\zeta$. We find that 76\% of the cases have a relative error smaller than 1\% and the mean relative error is 0.77\%. The  histogram of the relative errors is shown  in Fig.~
\ref{fig:zeta-hist} (left panel).  

\subsection{Observational Constraints}
\label{Sec:eor}

The parameter space outlined above is constrained by the available
observations of the EoR. 
In addition to the limits on $\tau$ from the
CMB experiments, we consider two other types of constraints when
developing the global signal emulator:
\begin{enumerate}
\item Stellar models indicate that for the extreme case of massive
  population III stars, $N_{\rm ion} =40,000$ \citep{Bromm:2001};
  therefore, we set an upper limit of $\zeta_{\rm max}=40,000 f_*$
  based on Eq.~\ref{Eq:zeta}. Hence, our first requirement for a
  parameter set to be valid is that $\zeta< \zeta_{\rm max}$.
\item Absorption seen in the spectra of high-redshift quasars measures
  the neutral fraction of the Universe
  \citep[e.g.,][]{Banados:2018}. A $2\sigma$ upper limit of $x_{\rm
    HI,~max} = 16\%$ on the neutral fraction at $z=5.9$
  ($\nu=205.85$ MHz) was derived from quasar absorption troughs
  \citep{McGreer:2015}. Our second requirement is, thus, $x_{\rm
    HI}(z=5.9) < 16\%$. In this paper we do not take into account the latest constraints from the Ly-$\alpha$ emitting galaxies \citep{Mason:2019} and high redshift quasars \citep[e.g.,][]{Banados:2018} as they became available when our paper was close to being completed \citep[however, see][]{Monsalve:2019}.  To incorporate the neutral fraction constraint in our modelling, we train a NN to predict at which frequency, denoted by
$\nu_{16\%}$, the neutral fraction reaches 16\% for the given set of
astrophysical parameters, $\nu_{16\%} = \nu\left(x_{\rm
  HI}=0.16\right)$. The reionization history is considered valid if this
frequency is lower than 205.85 MHz (i.e., the redshift is higher than
5.9). Because for many cases $x_{\rm HI}(z=5.9)$ is zero, $\nu_{16\%}$
can be more easily inferred with high accuracy than the neutral
fraction at $z=5.9$. Technical details of this NN are discussed in Section \ref{Sec:NNtrained}.
The performance of the NN in predicting
$\nu_{16\%}$ is evaluated in Figure \ref{fig:zeta-hist} (right panel) where we show
the histogram of relative errors. We find that the mean relative error
is 0.47\% and 92\% of cases have a relative error smaller than 1\%.
\end{enumerate}
 
\begin{figure*}
	\centering
	\includegraphics[width=2.7in]{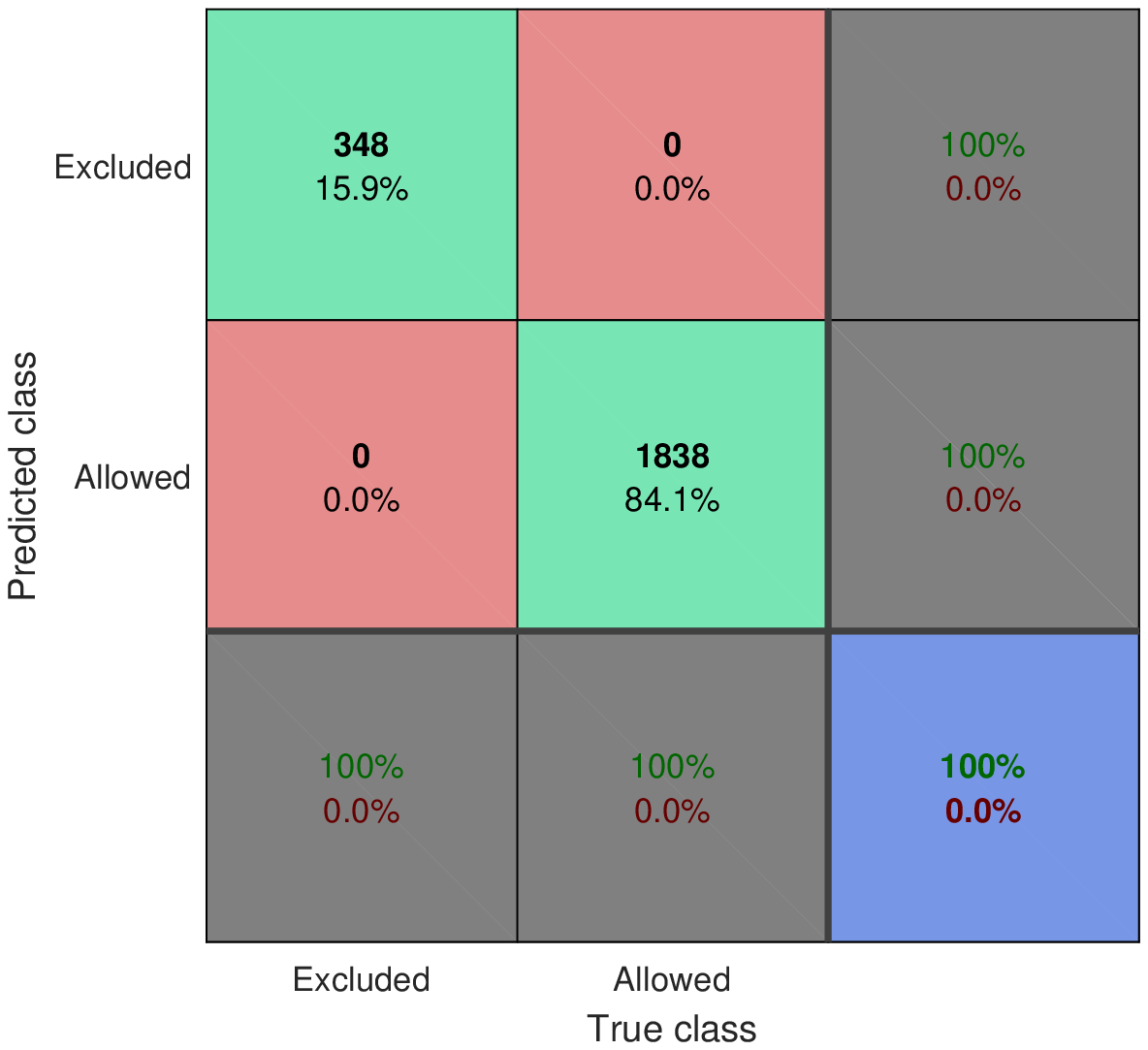}\hspace{0.4in}
	\includegraphics[width=2.7in]{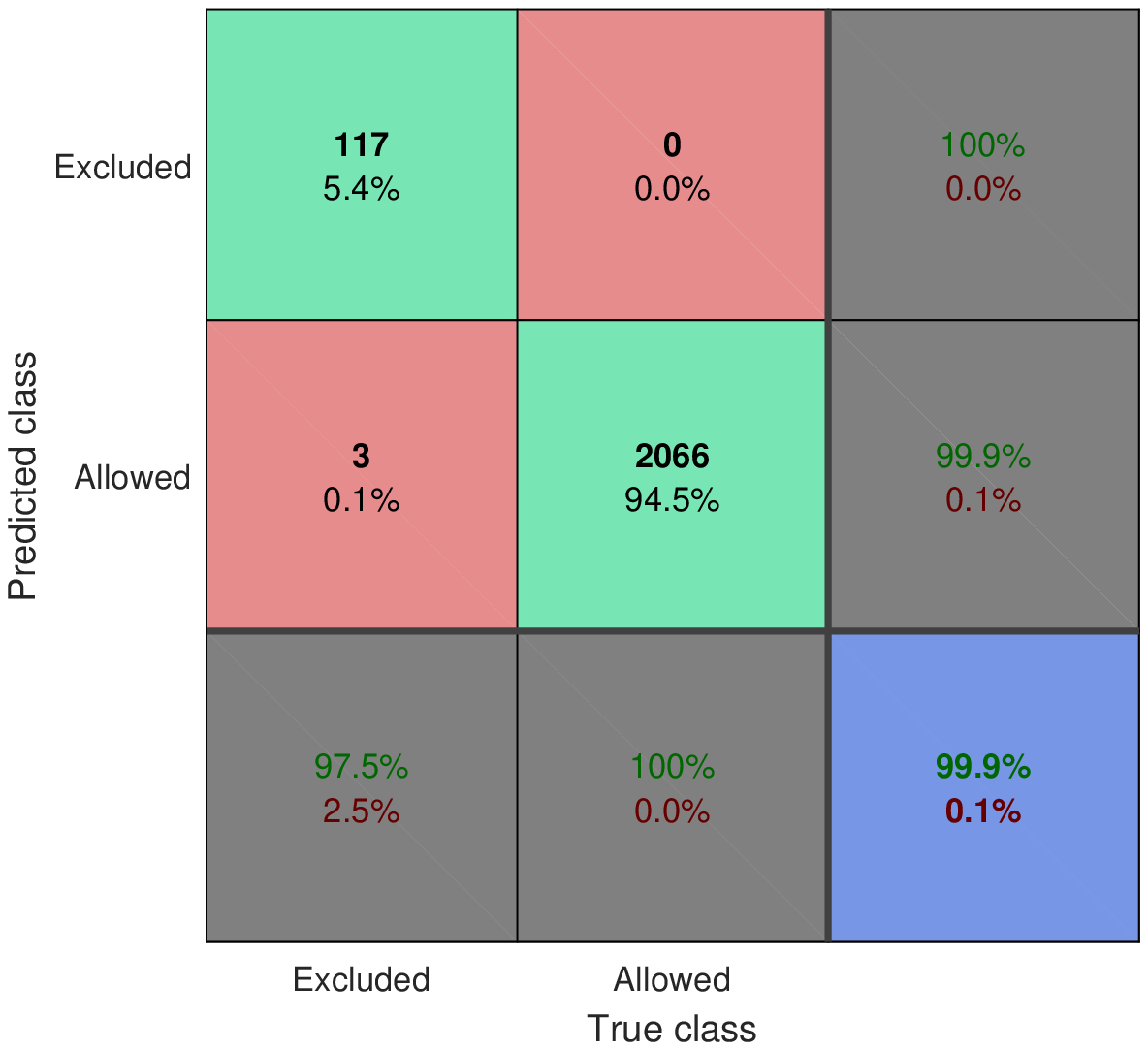}
	\caption{Left panel: The confusion matrix of the exclusion
          process $\zeta>\zeta_{\rm max}$. Right panel: The confusion
          matrix of the exclusion criterion $\nu_{16\%}>205.85$
          MHz. The structure of each confusion matrix is as follows:
          Each matrix has 9 fields with the green squares showing the
          number of excluded and allowed cases which were correctly
          classified, red squares showing the number of
          miss-classified cases, grey showing the percentage of the
          correct predictions for each row/column, and blue showing
          the total accuracy defined as the ratio of the total number
          of correctly identified cases (both excluded and allowed) to
          the total number of considered cases.  The classification is done correctly in 100\% of
cases for $\zeta$, and in 99.9\% of cases for $\nu_{16\%}$.}
	\label{fig:conf1}
\end{figure*}

As a part of the global signal emulator, described
in detail in Section \ref{Sec:Interp}, the code checks whether or not an input
parameter set renders a valid EoR history, i.e., given the generated
values of $\zeta$ and $\nu_{16\%}$, whether $\zeta<\zeta_{\rm max}$
and $\nu_{16\%}<205.85$ MHz. 
The success/failure rates of the
validation process is summarized in the form of confusion matrices
shown in Figure \ref{fig:conf1}. Out of 2186 tested cases, 348 are
excluded based on their values of $\zeta$ and 117 are excluded based
on the values of $\nu_{16\%}$ (22 overlap, i.e., are inconsistent with
either constraint). The classification is done correctly in 100\% of
cases for $\zeta$, and in 99.9\% of cases for $\nu_{16\%}$.
 
\begin{figure}
	\centering
	\includegraphics[width=2.8in]{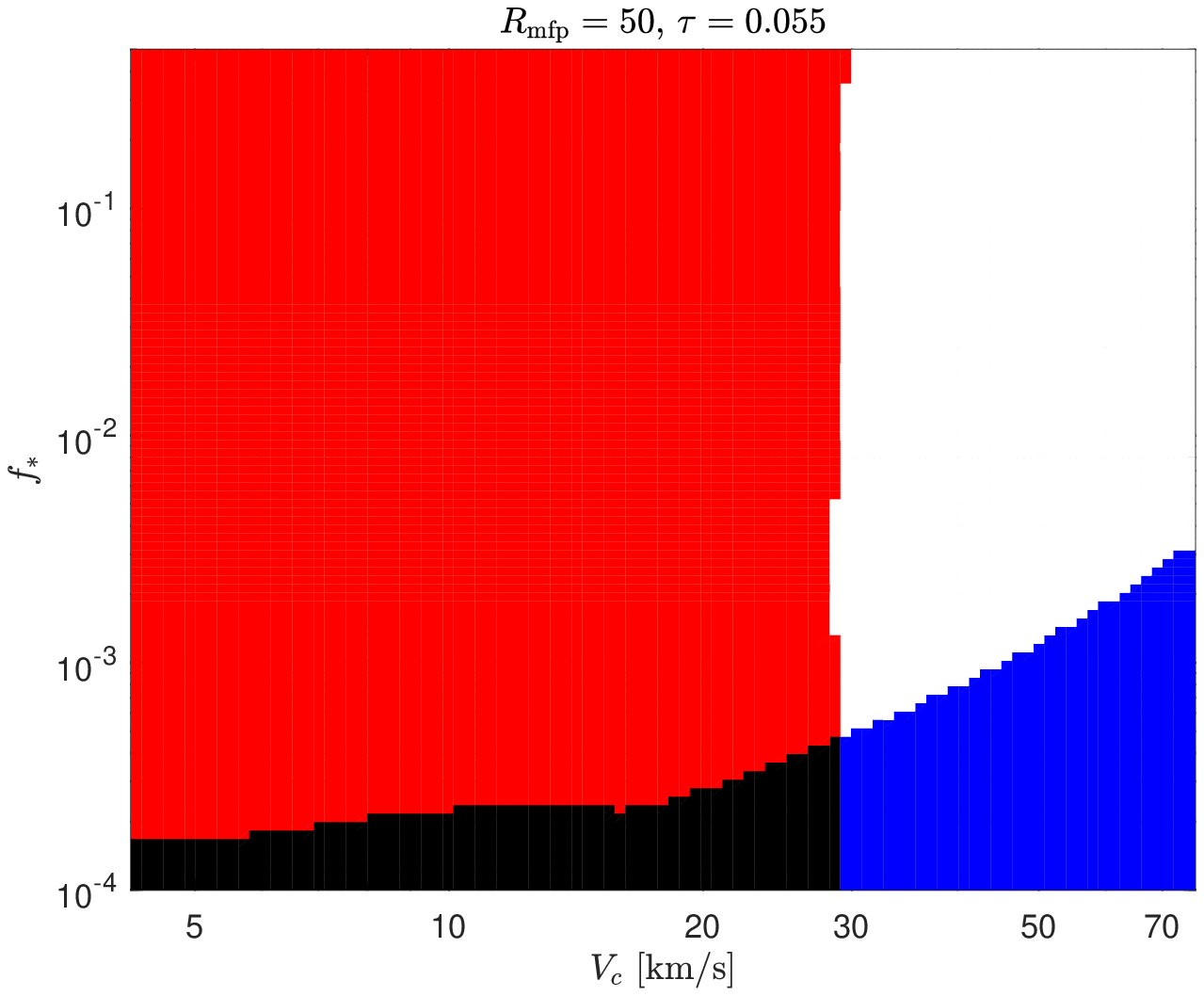}	
\hspace{0.2in}
	\includegraphics[width=2.8in]{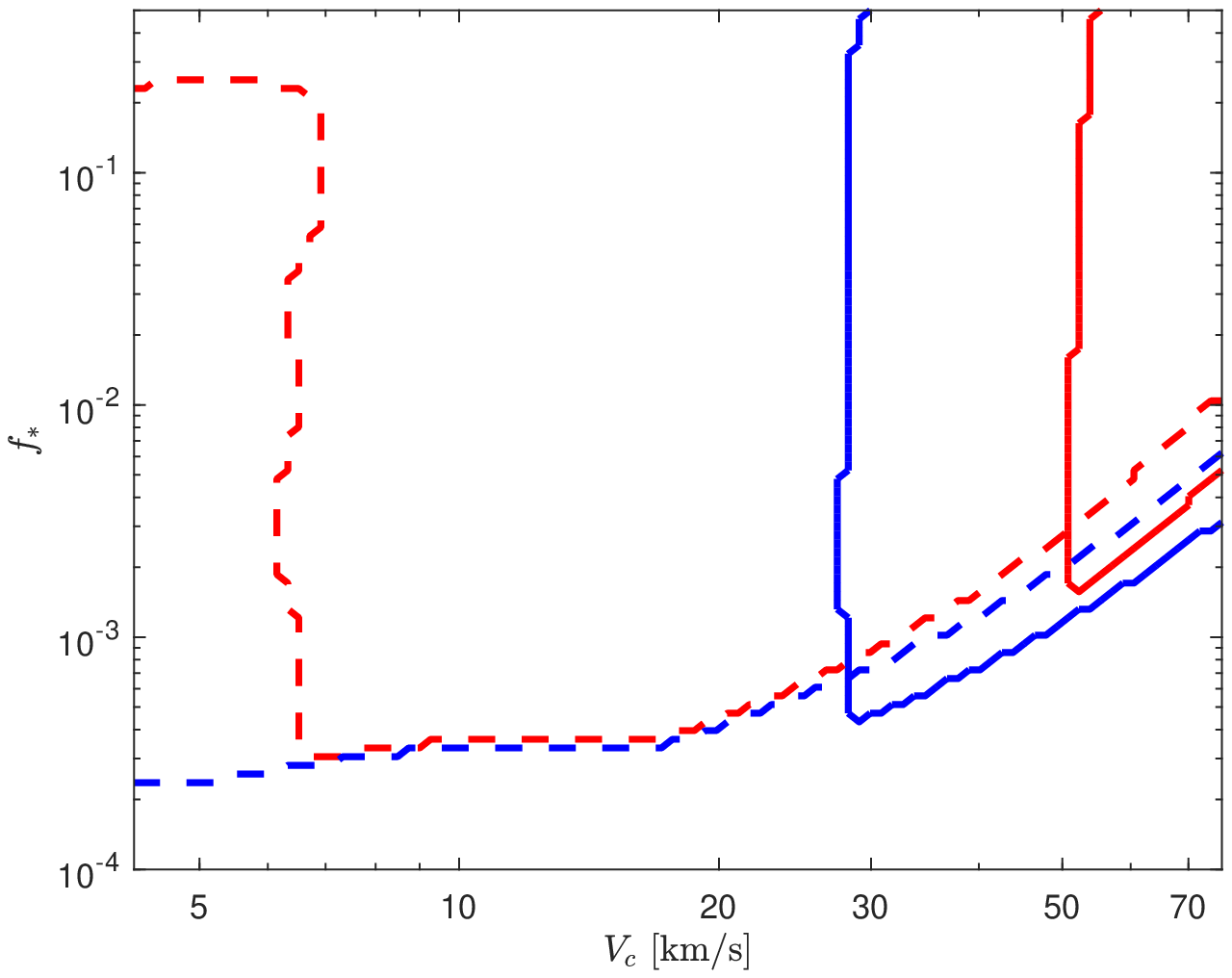}	
		\hspace{0.2in}
	\includegraphics[width=2.8in]{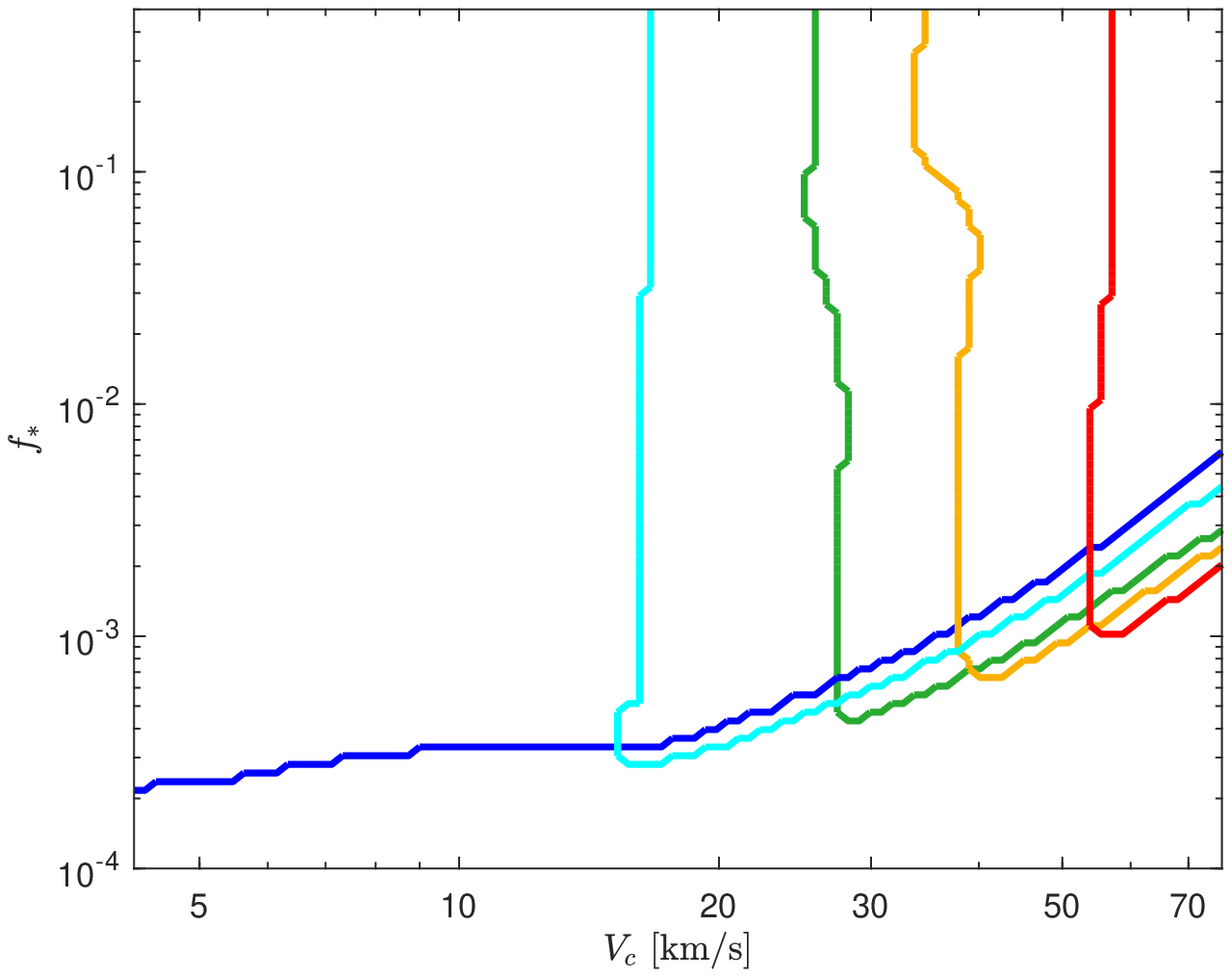}
	\caption{ Constraints on $V_c$ and $f_*$ imposed by the constrained reionization history. 
	Top: We show  allowed (white) and excluded (blue for
          $\zeta>\zeta_{\rm max}$, red for $x_{\rm HI}(z = 5.9)>16\%$,
          black for both) regions in the $f_*-V_c$ plane for $\tau =
          0.055$ and $R_{\rm mfp} =50$ Mpc. Also assumed are $f_X=1$,
          $\alpha=1.5$, and $\nu_{\rm min}=0.2$~keV.  Middle panel: The total exclusion contours (the excluded regions are under and to the left of the curves) for
          $\tau=0.055$ (solid lines), $\tau=0.064$ (dashed), $R_{\rm
            mfp} =10$ Mpc (red) and $R_{\rm mfp} =50$ Mpc (blue). The
          same X-ray parameters were
          assumed. Bottom panel: The total exclusion contours are shown
          for $\tau=0.049$ (red), 0.052 (orange), 0.055 (green), 0.060
          (cyan) and 0.064 (blue); these are the regions that are
          excluded for all values of $R_{\rm mfp}$ and X-ray
          parameters that we consider (i.e., these are {\em not}\/
          averaged over those parameter regions). The highest $\tau$
          which is completely excluded within our parameter space is
          $\tau=0.046$. }
	\label{fig:lims}
\end{figure}

The above-mentioned constraints on $\zeta$ and $\nu_{16\%}$ condition
the entire parameter space, because the reionization history depends on several astrophysical
parameters simultaneously.  Top panel of Figure \ref{fig:lims} illustrates the
mapping between the EoR constraints and the allowed regions in the
$f_* - V_c$ plane for the specific choice of the EoR parameters,
$\tau=0.055$ and $R_{\rm mfp}=50$ Mpc.  For each combination of $f*$ and $V_c$ we use the trained NNs to check whether the reionization history is valid or not. In the figure, the area
where the two exclusion criteria overlap is painted in black, the excluded region with $\zeta>\zeta_{\rm max}$ is shown in blue,
while the region with $x_{\rm HI}(z = 5.9)>16\%$ is red. The white regions have valid reionization histories.

The shape of the excluded and allowed regions is easy to
understand. Consider first the requirement $\zeta<\zeta_{\rm
  max}$. For given values of $\tau$ and $R_{\rm mfp}$ (as well as the
fixed heating parameters of $f_X=1$, $\alpha=1.5$ and $\nu_{\rm
  min}=0.2$ keV), models with a low star formation efficiency require
high values of $\zeta$ that exceed the upper limit. Therefore, cases
with low $f_*$ are excluded. Now, the lower the value of $V_c$ is, the
more star forming halos there are, making it easier to reionize (and
match the required value of $\tau$) without needing to exceed
$\zeta_{\rm max}$. Therefore, the maximum excluded $f_*$ is a
monotonically growing function of $V_c$. This function grows rapidly
at the highest $V_c$ due to the exponential dependence of the halo
abundance on $V_c$, while it changes slowly at $V_c<16.5$ km s$^{-1}$
because in this mass range the number of stars at a given $f_*$ is
regulated by the LW feedback by the time of the bulk of cosmic
reionization.
 
Consider the second requirement, $x_{\rm HI}(z = 5.9)<16\%$. Since
$\tau$ has fixed the average timing of reionization, the neutral
fraction constraint rules out cases with low $V_c$, since those are
characterized by a more gradual evolution of reionization and, thus, a
higher remaining neutral fraction at $z=5.9$, regardless of the values
of the other parameters. Therefore, this requirement rules out the
left portion of the $f_*-V_c$ plane (for a fixed $R_{\rm mfp}$).

The exclusion contours for several choices of $\tau$ and $R_{\rm mfp}$
at the fixed values of the heating parameters are shown in the  middle
panel of Figure \ref{fig:lims}, with the previously examined
(reference) case of $\tau = 0.055$ and $R_{\rm mfp} =50$ Mpc shown in
solid blue. A lower $R_{\rm mfp}$ implies a more gradual end to
reionization (thus raising the residual $x_{\rm HI}$ at $z=5.9$) since
sources then cannot contribute ionizing photons beyond this shorter
distance. A higher $\tau$ moves the bulk of reionization towards
higher redshifts, making it more compatible with the observational
constraint at the fixed redshift of 5.9. With a lower $R_{\rm mfp}$ of
10 Mpc (solid red), the excluded area is larger with a lowest allowed
value of $V_c=52$ km s$^{-1}$ ($\sim 28\times M^{\rm atomic}_{\rm
  min}$), compared to 29~km s$^{-1}$ ($\sim 5\times M^{\rm
  atomic}_{\rm min}$) for the reference case. Increasing $\tau$ (the
dashed lines correspond to $\tau = 0.064$) allows a wider range of
$V_c$. In that case, if $R_{\rm mfp}=10$ Mpc (dashed red) then only
$V_c<6.3$ km s$^{-1}$ is excluded, and even that is only if
$f_*\lesssim 0.25$. At the highest star formation efficiencies,
partial ionization by X-rays becomes significant, speeding up the
process of reionization.

After searching over the full range of $R_{\rm mfp}$ and the heating
parameters, we show the absolutely excluded regions for various values
of $\tau$ in the bottom panel of Fig.~\ref{fig:lims}; i.e., these are
regions that are always excluded, there is no averaging here. We found
a lower limit for the optical depth of $\tau=0.046$. For the best-fit
{\it Planck} value of $\tau=0.055$ we found lower limits on the
circular velocity of $V_c \sim 26$ km s$^{-1}$ ($\sim 4 \times M^{\rm
  atomic}_{\rm min}$) and on the star formation efficiency of $f_*
\sim 0.0004$. However, for $\tau = 0.064$ ($1\sigma$ away from the
best-fit \textit{Planck} measurement) no values of $V_c$ are excluded
and the absolute minimum on the star formation efficiency is $f_* \sim
0.0002$.

\subsection{Dataset}
Using the modeling outlined above we created a dataset of 29,641
global 21-cm signals that cover a very wide range of possible values
of the seven astrophysical parameters, $V_c = 4.2-100$ km s$^{-1}$,
$f_* = 0.0001-0.50$, $\alpha = 1 - 1.5$, $\nu_{\rm min} = 0.1-3$ keV,
$f_X =0-1000$, $\tau = 0.04-0.2$, $R_{\rm mfp} = 10-50$ Mpc, and
verifying whether or not the ionization history complies with the EoR
constraints (subsection~\ref{Sec:eor}). The sampling of the parameter
space was done randomly with uniform priors on
$\log_{10}\left(V_c\right)$, $\log_{10}\left(f_s\right)$,
$\log_{10}\left(f_X\right)$, $R_{\rm mfp}$ and $\tau$. The SED was
randomly chosen with $\alpha =$ 1, 1.3 or 1.5, and $\nu_{\rm min}=$
0.1, 0.2, 1 or 3 keV. The 21-cm spectra are created over the redshift range $z = 5-50$ and are sampled at $\Delta z = 0.1$. The set of models was (randomly) split into the training and testing sets. The parameters of the testing set are restricted to $V_c = 4.2-76.5$ km s$^{-1}$,
$f_* = 0.0001-0.50$, $\alpha = 1 - 1.5$, $\nu_{\rm min} = 0.1-3$ keV,
$f_X =0-10$, $\tau = 0.055-0.1$, $R_{\rm mfp} = 10-50$ Mpc over which ranges the performance of {\sc 21cmGEM} was optimized.
The training and testing datasets are  available online at  \url{https://www.ast.cam.ac.uk/~afialkov/Publications.html}.

\section{Consistency Relations}
\label{Sec:Data}

\citet{Cohen:2017} derived universal relations between astrophysical
quantities (such as the heating rate, $\epsilon_X$, and the intensity
of the Ly$\alpha$ background, $J_\alpha$), and the three key points of
the global signal, including the high-z maximum at the redshift
labeled $z^{\rm hi}_{\rm max}$ (at matching frequency $\nu^{\rm
  hi}_{\rm max}$) and the brightness temperature $T^{\rm hi}_{\rm
  max}$; the absorption trough located at $z_{\rm min}$ (or $\nu_{\rm
  min}$) and reaching $T_{\rm min}$; and the low-z maximum at $z^{\rm
  lo}_{\rm max}$ (or $\nu^{\rm lo}_{\rm max}$) with $T^{\rm lo}_{\rm
  max}$. That work was based on a dataset of 193 signals generated
using a 5-parameter model ($V_c$, $f_*$, $\tau$, $f_X$ and either a
hard or soft X-ray SED) with the parameters sampled on a grid
\citep[see][ for a detailed description of the sampling]{Cohen:2017}.
Here we verify the validity of the above-mentioned relations in the
context of our extended 7-parameter model and using a sub-set of 1948
randomly drawn combinations of the parameters.
We find a good general agreement between this work and   the previous study. However, compared to the previous study, we find significantly larger scatter owing to the larger explored astrophysical parameter space. 

\begin{figure}
	\centering
	\includegraphics[width=2.8in]{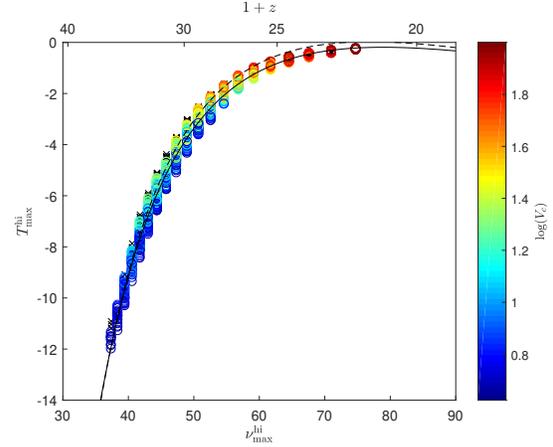}
	\caption{Brightness temperature at the high-redshift maximum point as a function of the observed
          frequency $\nu^{\rm hi}_{\rm max} = 1420~{\rm MHz}/(1+z^{\rm hi}_{\rm max})$. The colors
          indicate the value of $V_c$ as indicated on the colorbar:
          dark blue corresponds to the lowest value of $V_c$ (4.2 km
          s$^{-1}$), and dark red corresponds to its highest value
          (76.5 km s$^{-1}$). Also shown is a fitting function
          (Eq.~\ref{eq:hmax}, solid) along with our older fit from
          Eq.~8 of \citet{Cohen:2017} (dashed) for comparison. Black
          $\times$'s show models that were excluded by our
          observational constraints. We observe a tight correlation between $T^{\rm hi}_{\rm max}$ and $\nu^{\rm hi}_{\rm max}$.}
	\label{fig:hmax}
\end{figure}

At the onset of Cosmic Dawn the 21-cm signal is driven by atomic
physics and the early process of Ly$\alpha$ coupling due to star
formation, which results in a close relation between $z^{\rm hi}_{\rm
  max}$ and $T^{\rm hi}_{\rm max}$ as shown in
Fig.~\ref{fig:hmax}. There is low scatter relative to a relation that
can be fitted with a quadratic function of the form
\begin{equation}
\label{eq:hmax}
T^{\rm hi}_{\rm max} = a\left(1+z^{\rm hi}_{\rm max} \right)^2 +b\left( 1+z^{\rm hi}_{\rm max}\right) + c ~.
\end{equation}
 Using the extended dataset we find a similar relation to the one
 reported by \citet{Cohen:2017} (Eq.\ 8 and Fig.\ 2 in that paper),
 with the best-fit parameters changed by 10-20\%. The new best-fit
 values are $[a,b,c]=[-0.02925,1.053,-9.667]$.

\begin{figure*}
	\centering
	\includegraphics[width=2.8in]{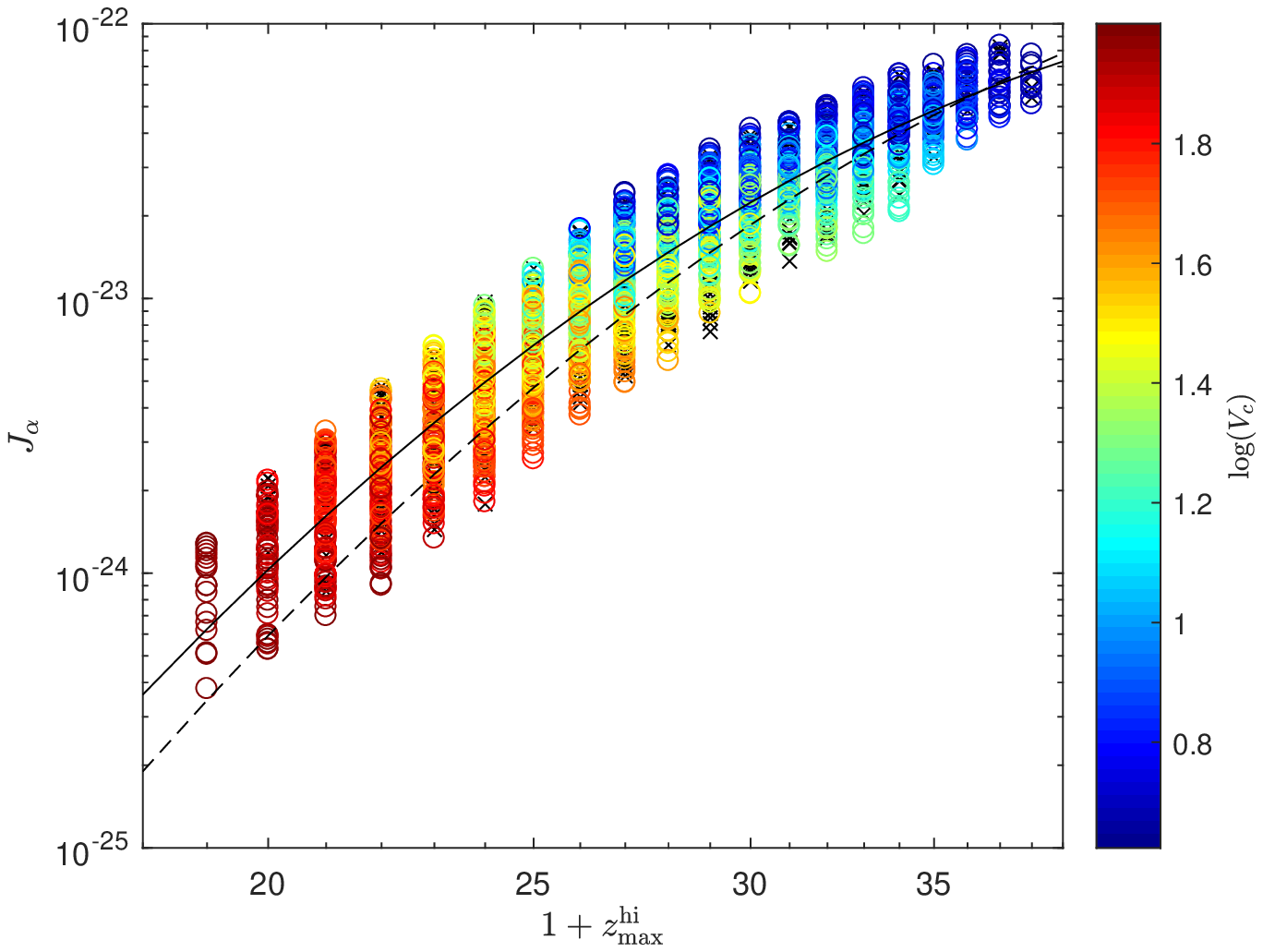}\hspace{0.2in}
	\includegraphics[width=2.8in]{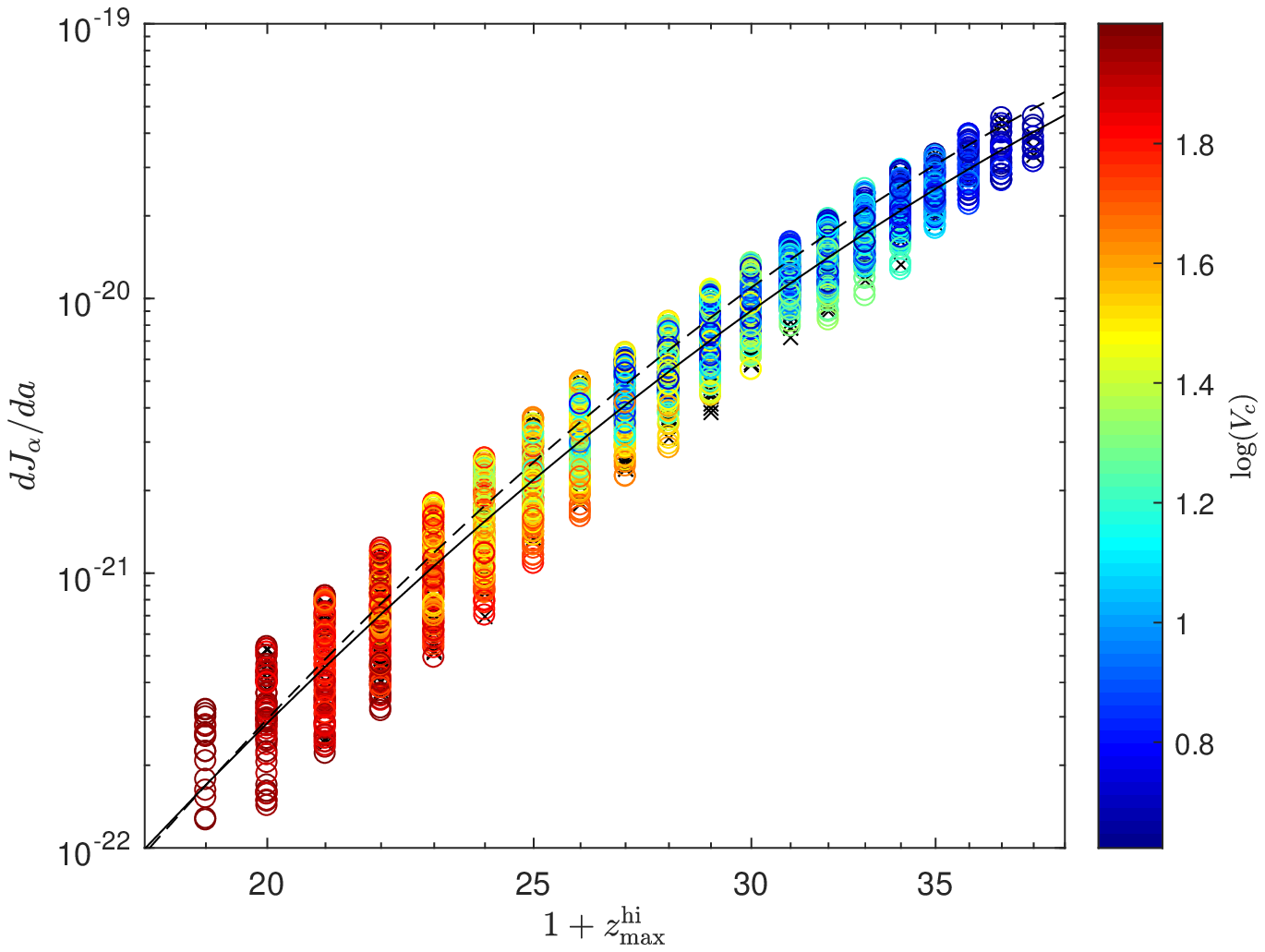}
	\caption{The Ly$\alpha$ intensity in units of erg s$^{-1}$
          cm$^{-2}$ Hz$^{-1}$ sr$^{-1}$ (left) and its derivative with
          respect to the scale factor (right) as a function of $z^{\rm
            hi}_{\rm max}$. The colors indicate the value of $V_c$ in
          accordance with the colorbar. Also shown are the fitting
          function for the present data set (solid, Eq.~\ref{eq:Ja}
          and Eq.~\ref{eq:dJa} on the left and right panels,
          respectively); the fits from \citet{Cohen:2017} (dashed,
          Eq.~9 and Eq.~10 on the left and right panels, respectively)
          are shown for comparison. Black $\times$'s show models that
          were excluded by our observational constraints.  The large scatter in $J_\alpha$ is a result of neutral hydrogen excitation by X-rays. }
	\label{fig:hmaxJA}
\end{figure*}

The value of $z^{\rm hi}_{\rm max}$ (and, hence, the value of the
brightness temperature at this redshift) directly depends on the
intensity of the Ly$\alpha$ background that drives the WF
coupling. Therefore, it is natural to expect that the intensity of the
Ly$\alpha$ background can be inferred from the high-redshift maximum
of the signal. Following \citet{Cohen:2017} [see their Eqs.~9 and 10
and Fig.~3], we examine the relationship between $z^{\rm hi}_{\rm
  max}$ and the mean Ly$\alpha$ intensity measured at this redshift,
as well as its derivative with respect to the scale factor $a =
1/(1+z)$, and show the new results in Fig.~\ref{fig:hmaxJA}. The best
fits to the new data are:
\begin{equation}
\label{eq:Ja}
\log(J_\alpha) = a_1 \log^2\left( 1+z^{\rm hi}_{\rm max}\right) +b_1
\log\left( 1+z^{\rm hi}_{\rm max}\right) +c_1 ~,
\end{equation}
and
\begin{equation}
\label{eq:dJa}
\log \left( \frac{J_\alpha}{da}\right) = a_2 \log^2\left( 1+z^{\rm
  hi}_{\rm max}\right) +b_2 \log\left( 1+z^{\rm hi}_{\rm max}\right)
+c_2 ~,
\end{equation}
where $[a_1,b_1,c_1]=[-10.64,37.15,-54.31]$ and
$[a_2,b_2,c_2]=[-7.851,30.34,-47.73]$.  On average we find a good
agreement between the new study and the results of
\citet{Cohen:2017}. However, the scatter in $J_\alpha$ is now
substantially larger due to the contribution of X-rays to the
Ly$\alpha$ background via X-ray excitation of neutral hydrogen.

\begin{figure*}
	\centering
	\includegraphics[width=2.8in]{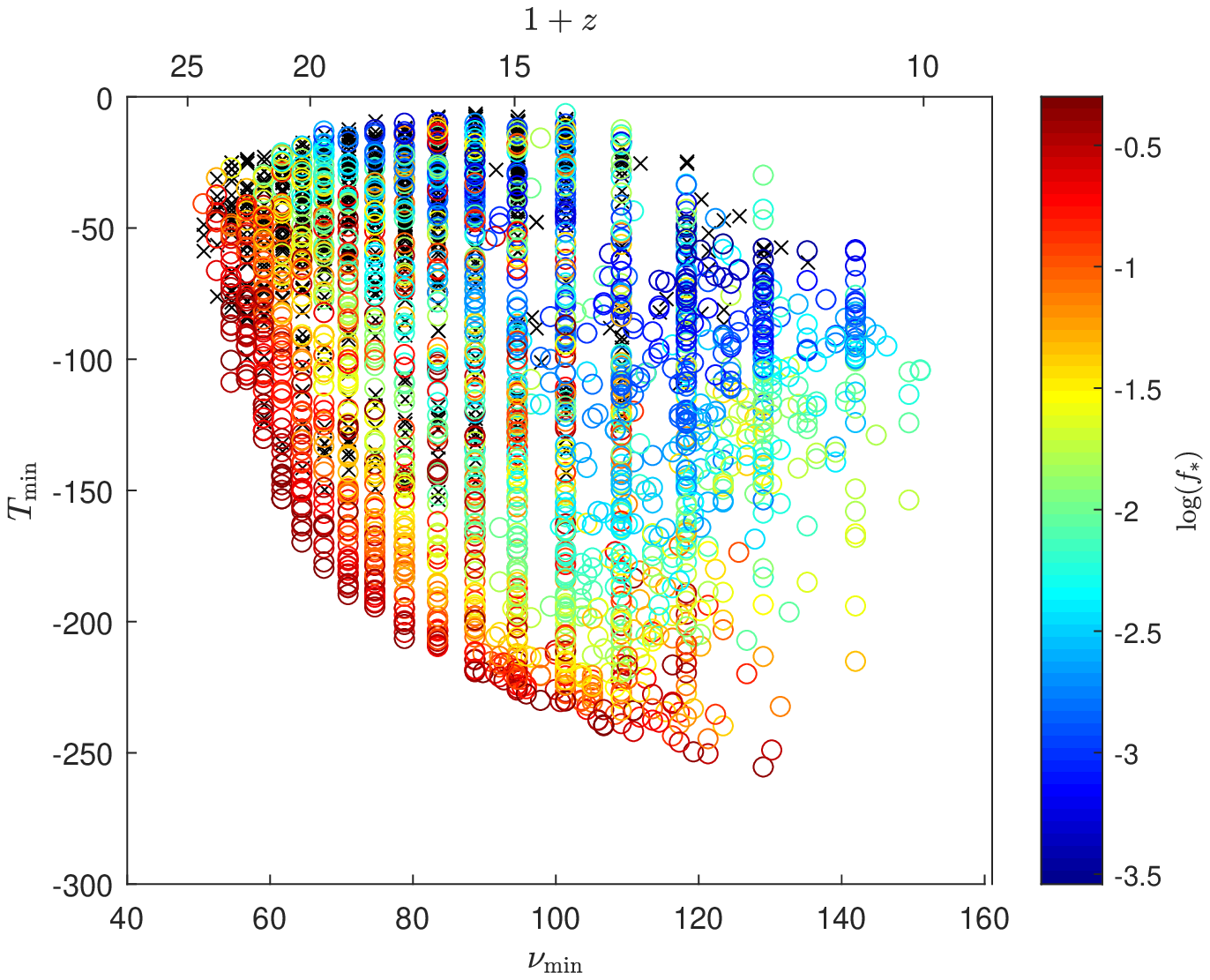}
		\hspace{0.2in}
	\includegraphics[width=2.8in]{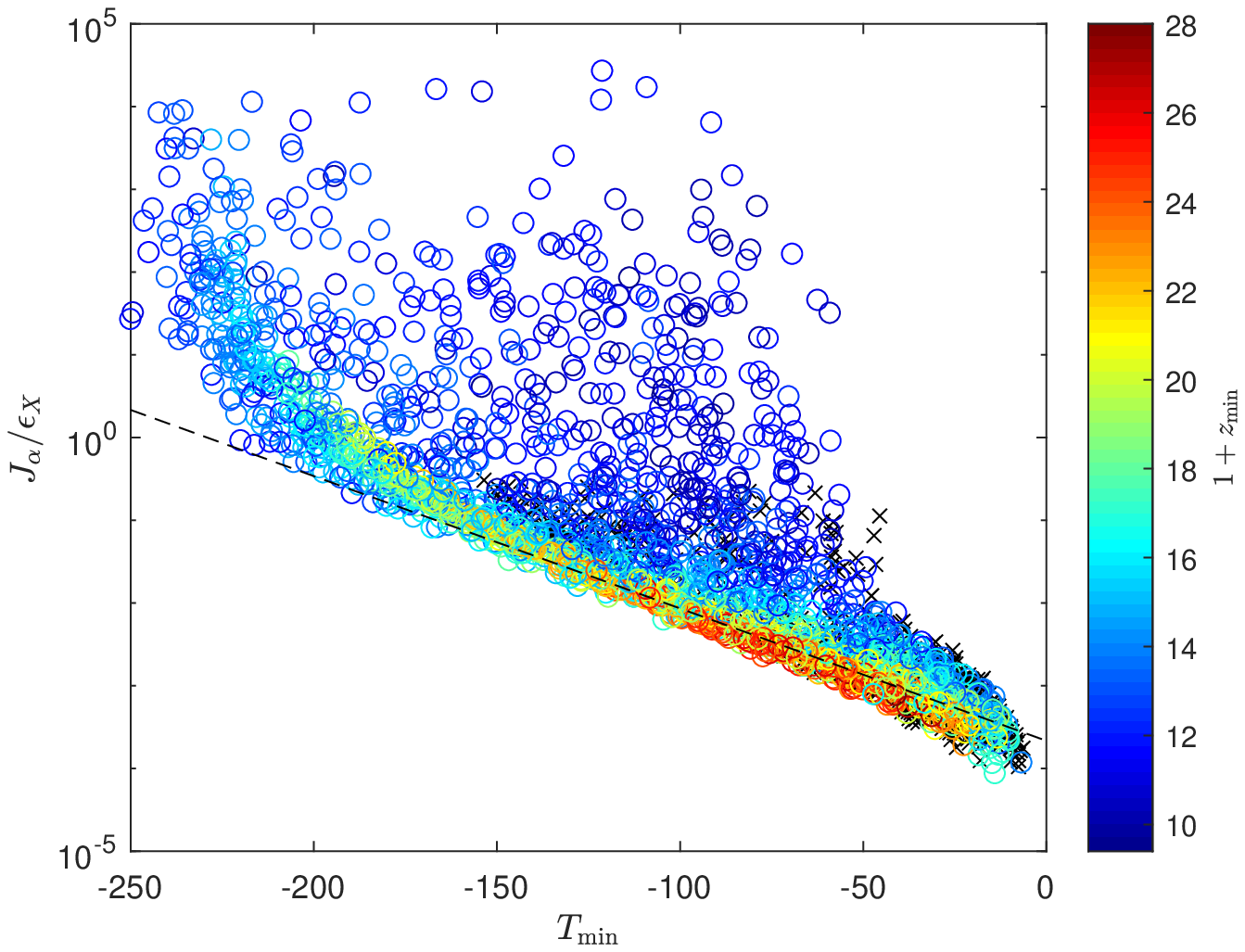}
	\caption{Left panel: Brightness temperature at the minimum
          point as a function of observed frequency of this point
          (bottom axis) or the equivalent one plus redshift (top
          axis). The colors indicate the value of the star formation
          efficiency (see the colorbar on the right).  High values of $f_*$ are needed for efficient WF coupling and deep absorption troughs. Right panel:
          The ratio between the Ly-$\alpha$ intensity (in units of erg
          s$^{-1}$ cm$^{-2}$ Hz$^{-1}$ sr$^{-1}$) and the X-ray
          heating rate (in units of eV s$^{-1}$ baryon$^{-1}$)
          measured at $z_{\rm min}$ as a function of the brightness
          temperature at the minimum point. The colors indicate the
          redshift of the minimum point (see the colorbar). Also shown
          is the fitting function from \citet{Cohen:2017} (dashed,
          Eq.~13). Black $\times$'s show models that were excluded by
          our observational constraints.  Large variation in the properties of X-ray sources explored in this work contributes to the larger  scatter in the $J_\alpha/\epsilon_X$ relation.}
	\label{fig:min}
\end{figure*}

The complexity of the Universe increases as the population of the
first heating sources forms. The location and the amplitude of the
absorption trough show a very large scatter \citep[left panel of
  Figure \ref{fig:min}, see also Fig.\ 4 of][]{Cohen:2017} due to the
dependence of the signal on both the parameters of heating and of star
formation. The latter regulates the strength of the WF coupling: for
an efficient WF coupling, $T_S$ is close to the kinetic temperature of
the gas (and the absorption trough is deeper); while for a very
inefficient coupling $T_S$ moves towards the temperature of the
background radiation (and the absorption trough is shallower). On the
other hand, the role of the X-ray sources is to heat up the gas: the
weaker the heating is,  the more time the Universe has to cool  down as a result of the
adiabatic expansion. Therefore, we get a lower limit given by the
strongest possible absorption in the case of a fully coupled,
adiabatically cooled gas:
\begin{equation}
\label{eq:min}
T_{\rm min}\geq 26.8\left(\frac{1+z_{\rm min}}{10} \right) ^{1/2}
\left(1- \frac{1+z_{\rm dec}}{1+z_{\rm min}} \right) \rm mK\ ,
\end{equation}
where $z_{\rm dec}=137$ \citep{Cohen:2017}.  The depth of the
absorption trough as a function of $\nu_{\rm min}$ is shown in the
left panel of Fig.~\ref{fig:min} color-coded as a function of $f_*$.

\citet{Cohen:2017} suggested that the ratio between the Ly-$\alpha$
intensity and the X-ray heating rate can be inferred from the value of
the brightness temperature at the minimum point. The right panel of
Figure~\ref{fig:min} shows that this does not entirely persist. The
larger variation in the properties of X-ray sources employed here
compared to what was implemented by \citet{Cohen:2017}, results in a
large scatter in the $J_\alpha/\epsilon_X$ relation. In particular,
models with very low $f_X$ (values that are unusually low compared
with low-redshift galaxies, but are still possible) break this
relation. As shown in the figure, this is the case only for models for
which the measured value of $z_{\rm min}$ would be low.

\begin{figure*}
	\centering
	\includegraphics[width=2.8in]{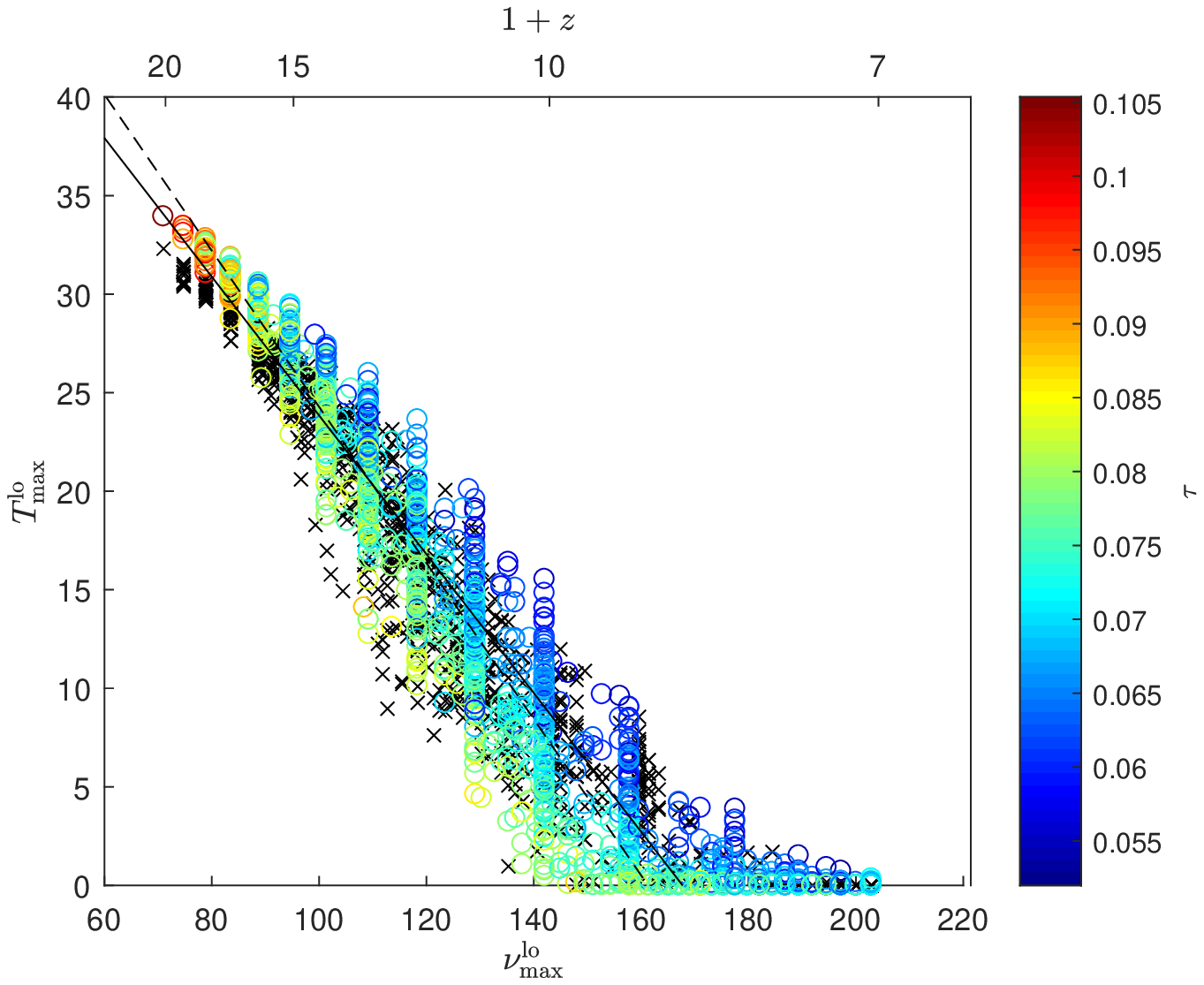}
	\hspace{0.2in}
	\includegraphics[width=2.8in]{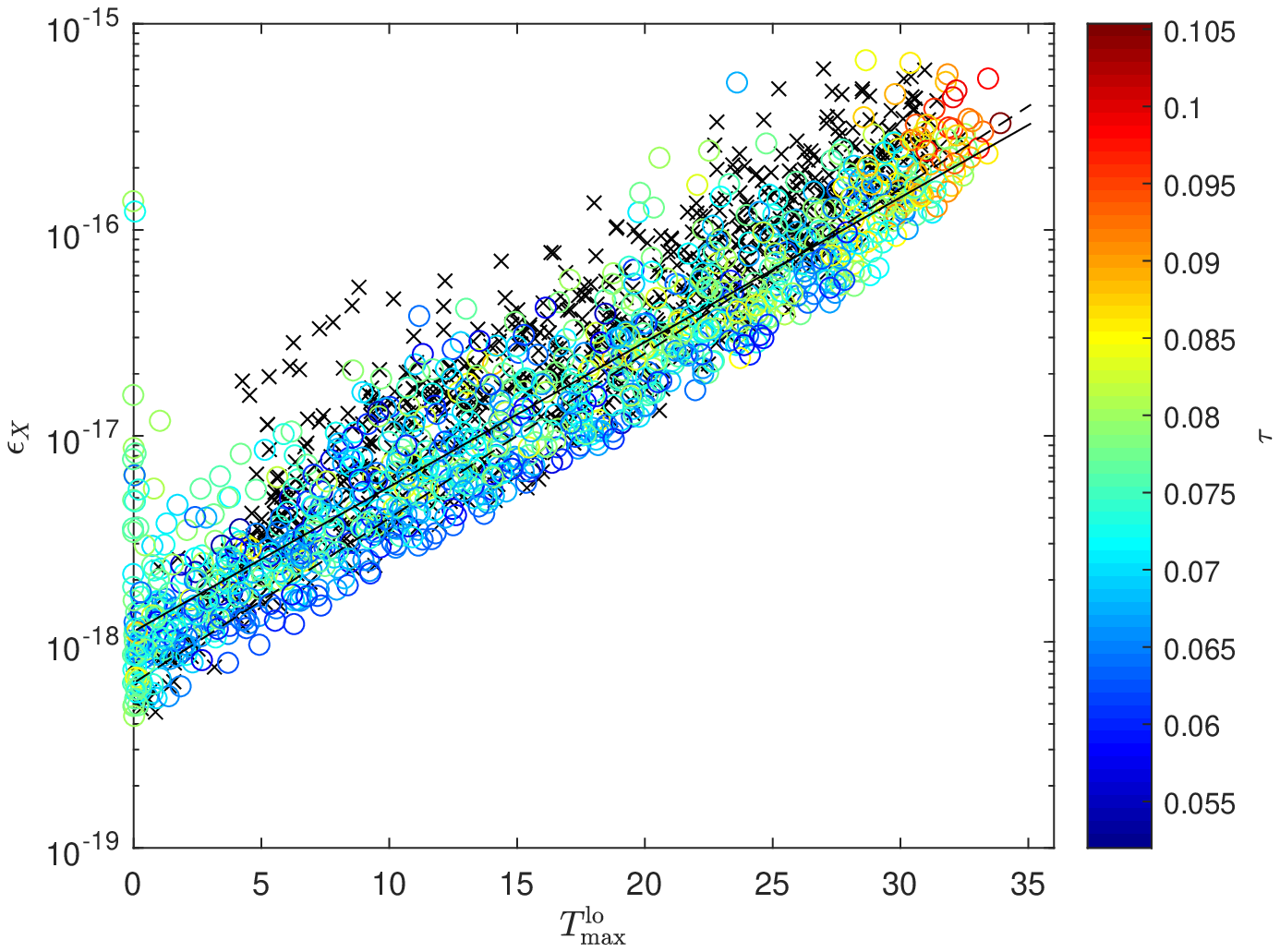}
	\caption{ Left panel: Brightness temperature as a function of
          observed frequency of the low-redshift maximum point. The
          colors indicate the CMB optical depth as is indicated on the
          colorbar. We also show the fitting function for the present
          data set (solid, Eq.~\ref{eq:Tl}) and the fit from
          \citet{Cohen:2017} (dashed, Eq.~15).  Owing to the constrained EoR history, the scatter in the $T^{\rm lo}_{\rm max}-z^{\rm lo}_{\rm max}$ relation is low. Right panel: X-ray
          heating rate (in units of eV s$^{-1}$ baryon$^{-1}$) as a
          function of the brightness temperature at the low-z maximum.
          The colors indicate the CMB optical depth as is indicated on
          the colorbar.  We also show the fitting function for the
          present data set (solid, Eq.~\ref{eq:minEPS}) and the fit
          from \citet{Cohen:2017} (dashed, Eq.~16). Black $\times$'s
          show models that were excluded by our observational
          constraints.  The nature of X-ray sources can be directly constrained from the measurement of $\epsilon_X$. }
	\label{fig:lmax}
\end{figure*}

Finally, we examine the emission peak of the 21-cm signal during the
EoR.  As was pointed out by \citet{Fialkov:2014b}, for a large part of
the astrophysical parameter space X-ray heating plays a major role in
the 21-cm signal during the EoR. Specifically, in cases of extremely
inefficient heating there is no transition of the 21-cm signal into
emission and the signal is seen in absorption throughout cosmic
history. Therefore, the location and the amplitude of the emission
feature depend not only on the EoR parameters but also on the heating
rate (as well as on the parameters of star formation). Because of the
complex dependence, one would expect to find a large scatter in the
values of ($z^{\rm lo}_{\rm max}$, $T^{\rm lo}_{\rm max}$). However,
as can be seen from the left panel of Figure \ref{fig:lmax} \citep[see
  also Eq.\ 15 and Fig.\ 7 of][]{Cohen:2017}, the scatter is
relatively low because the EoR history is significantly constrained by
current observations (Section \ref{Sec:eor}). The location and
amplitude of the emission peak for the present dataset are in good
agreement with our previous results. The relation can be fitted with:
\begin{equation}
\label{eq:Tl}
T^{\rm lo}_{\rm max} = \begin{cases}
a\frac{1}{1+z^{\rm lo}_{\rm max}}+b &\text{if $1+z^{\rm lo}_{\rm max}>\frac{-a}{b}$}\\
0 &\text{otherwise}
\end{cases}
\end{equation}
where $[a,b]=[-500.1,59.05]$.

 The right panel of Fig.~\ref{fig:lmax} shows the relation between the
 amplitude of the emission feature and the heating rate at
 $z^{\rm lo}_{\rm max}$, which can be fitted with
	\begin{equation}
\label{eq:minEPS}
\log\left( \epsilon_X\right)  = a T^{\rm lo}_{\rm max} + b,
\end{equation}
with $[a,b]=[0.07026,-17.95]$. While there is significant scatter,
this dependence can be used to constrain the properties of X-ray
sources directly from the measurement of the global 21-cm signal.

\section{{\sc 21cmGEM}: The Global Signal Emulator}
\label{Sec:Interp}

The main product of this work is the global signal emulator which,
given a set of 7 input astrophysical parameters, outputs a realization
of the global 21-cm signal sampled at $\Delta z = 0.1$ over the redshift range $z = 5-50$. In addition to the 21-cm spectrum, {\sc 21cmGEM} outputs frequencies at which the neutral fraction is 0.16\% and 11\% along with  values of the neutral fraction at $z=5.9$, $z=7.08$ and $z=7.54$. The values of the neutral fraction can be compared to the observational constraints on the reionization history at these redshifts: \citet{McGreer:2015} published the upper limit $\bar{x}_{\rm HI} < 0.06 + 0.05$ (at 68\% confidence)  at $z=5.9$, 
\citet{Greig:2017} find $\bar{x}_{\rm HI} = 0.40^{+0.21}_{-0.19} $ (68\%) from the damping wing analysis of a quasar at $z=7.08$; 
while \citet{Banados:2018} find  $\bar{x}_{\rm HI} =0.65^{+0.15}_{-0.32} $ (68\%) at $z=7.54$ using the  spectrum of ULASJ1342+0928, the
highest-redshift quasar detected so far.  This auxiliary information can be used to apply external constraints to the models \citep[see][]{Monsalve:2019}. 

Designed to detect features in the global
21-cm signal, the total-power experiments are very sensitive to steps
and wiggles in the data. To avoid spurious apparent detections, the
smoothness of the mock 21-cm signal over the entire observed frequency
band is one of the major requirements. Predicting the signal in each
frequency bin separately (as is done with the power spectrum
emulators) is not sufficient as it leads to discontinuities in the
spectrum. Instead, our approach here is to decompose the signals onto
a new basis of smooth functions that span the entire simulated
dataset. Principal Component Analysis  \citep[PCA,][]{Pearson:1901} is employed to find the
basis of such functions. Dividing the entire database into training
and testing sets, we train neural networks to predict the PCA
coefficients, along with the key points of the global signal, for any
input set of astrophysical parameters.  This information is then used
to generate the output 21-cm signal. The main steps of the emulation
process, as well as details of the training and optimization of the
algorithm, are described in the rest of this section.

\subsection{Design}
\subsubsection{Classification}
\label{Sec:Class}

As our parameter study shows, all the analyzed global signals have a
universal shape featuring a high redshift maximum and an absorption
trough (Note: for a reminder of the overall shape of the global 21-cm
signal, see the examples shown in Fig.~\ref{fig:ex}). The only
non-universal feature is the emission signal during the EoR which is
either present (we refer to this type of signal as {\it positive}) or
not ({\it negative} signals) depending on the astrophysics. Because of
this fundamental difference in the shape of the signals, our algorithm
is two-fold and treats the two types of signals separately.  The
classification into positive and negative cases is an essential part
of the training and the prediction processes.  If the signal is
positive, it has four key points: the high-redshift maximum,
absorption trough, low-redshift maximum and the redshift of complete
reionization ($\nu^{\rm hi}_{\rm max}$, $\nu_{\rm min}$, $\nu^{\rm
  lo}_{\rm max}$, and $\nu_{\rm reion}$, respectively). A negative
signal has only 3 key points ($\nu^{\rm hi}_{\rm max}$, $\nu_{\rm
  min}$, and $\nu_{\rm reion}$, respectively).  The key points divide
each positive (negative) case into 3 (2) segments. As we detail in the
next subsection, each segment is analyzed separately using the PCA.

\begin{figure}
	\centering
	\includegraphics[width=2.8in]{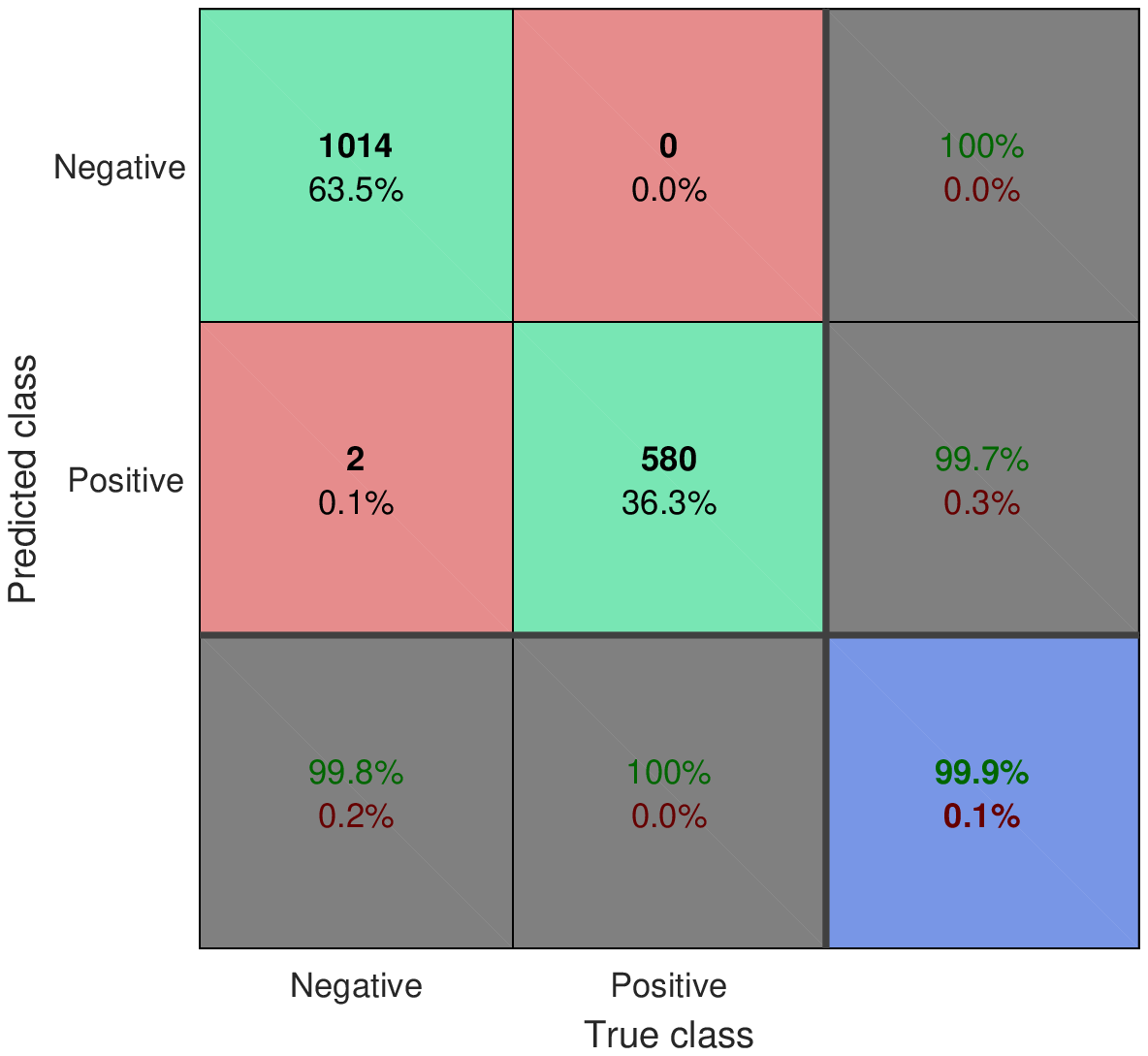}
	\caption{Confusion matrix of the classifier. The green squares
          show the number of correctly classified positive and
          negative cases, the red squares show the number of
          mis-classified cases, grey shows the percentage of the
          correct predictions for each row/column, and blue shows the
          total accuracy defined as the ratio of the total number of
          correctly identified cases (both positive and negative) to
          the total number of considered cases.  The
classification is correct  in 99.9\% of cases.}
	\label{fig:conf2}
\end{figure}

The {\it bagged trees} algorithm \citep{Breiman:1994, Loh:1997} was used to determine whether a case
is negative or positive. This algorithm fits many decision trees, each
time using a different subset of the training set, and the decision is
made by voting. After optimization we chose to use bagged trees with
30 tree learners and tree size chosen using 5-fold
cross-validation. The classification was tested on 1014 negative and
580 positive cases. We first tested the accuracy of the classification
process against each of the test cases and visually compared the
results to assess the performance. Knowing the location of the
emission feature (low-$z$ maximum point) compared to the timing of the
other key points helps to improve the quality of the
classification\footnote{ We found that the original algorithm misclassified 5.5\% of negative cases as positive. 
For almost all of these cases, at the output of the algorithm the order of the predicted redshifts of the key points was wrong  (e.g.,  the redshift of the low-z maximum point was predicted to be higher than the redshift of the absorption trough, which is unphysical). We used the ill-ordered key points as a diagnostic and for such cases changed the classification of the model  from positive to negative.  After this procedure, the
algorithm returned the correct answer in 99.9\% of cases.}. The success rate of the algorithm is 99.9\% as is demonstrated by the confusion matrix (Figure~\ref{fig:conf2}). Note that for
this test we only used cases with $T^{\rm lo}_{\rm max}>0.2$ mK. This
is because cases with a lower (but still positive) emission peak are
really neither positive nor negative, and mis-classification in this
case does not lead to an inaccurate prediction of the 21-cm signal
itself.

\subsubsection{PCA}
 \label{Sec:PCA}

The core of our emulator is PCA which, given a database, finds an
orthogonal basis that spans the data. Eigenfunctions (or eigenvectors)
of this basis are smooth functions found using the covariance matrix
of the data; while eigenvalues are a measure of the variance of the
data along each particular eigenvector. The basis is constructed so
that the first principal component (the eigenvector with the largest
eigenvalue) has the largest possible variance, the component with the
second greatest variance is the second principal component, and so on.
Using the basis of smooth functions to represent the 21-cm signal
guarantees the smoothness of the outcome.

The astrophysical key points divide each positive/negative signal into
3/2 distinct frequency segments. For a positive signal the segments
are $s_1 \in [\nu^{\rm hi}_{\rm max}, \nu_{\rm min}]$, $s_2^p \in
[\nu_{\rm min}, \nu^{\rm lo}_{\rm max}]$, and $s_3^p \in [ \nu^{\rm
    lo}_{\rm max}, \nu_{\rm reion}]$; while for a negative signal the
segments are $s_1 \in [\nu^{\rm hi}_{\rm max}, \nu_{\rm min}]$ and
$s_2^n \in [\nu_{\rm min}, \nu_{\rm reion}]$. We find it best to split
the data into these segments and analyze them separately. Note that,
because the first segment, $s_1$, is defined identically for both
positive and negative cases, over $s_1$ all signals are analyzed
together; while over other segments the positive and negative cases
are treated separately. In order to uniformly normalize the signals
within each segment of the data, we perform a coordinate transformation
into a new coordinate system $x_s, y_s$ in which each signal varies in
the range $x_s\in[0,1]$ and $y_s \in [-1,1]$. For instance, on $s_1$
the following coordinate transformation from the $\nu-T_{21}$ plane to
the $x_s-y_s$ plane is performed:
\begin{equation}
\label{eq:xs}
x_s=\frac{\nu-\nu^{\rm hi}_{\rm max}}{\nu_{\rm min}-\nu^{\rm hi}_{\rm max}},~
y_s=\frac{T-T^{\rm hi}_{\rm max}}{T^{\rm hi}_{\rm max}-T_{\rm min}}\ .
\end{equation}
In other words, for both the negative and positive signals $s_1$ is
chosen so that $(\nu^{\rm hi}_{\rm max},T^{\rm hi}_{\rm max})$ is mapped to
$(x_s,y_s)= (0,0)$, and the minimum point $(\nu_{\rm min},T_{\rm
  min})$ is mapped to $(x_s,y_s)= (1,-1)$. Each re-normalized segment
is separately analyzed using PCA.

\begin{figure}
	\centering
	\includegraphics[width=3.2in]{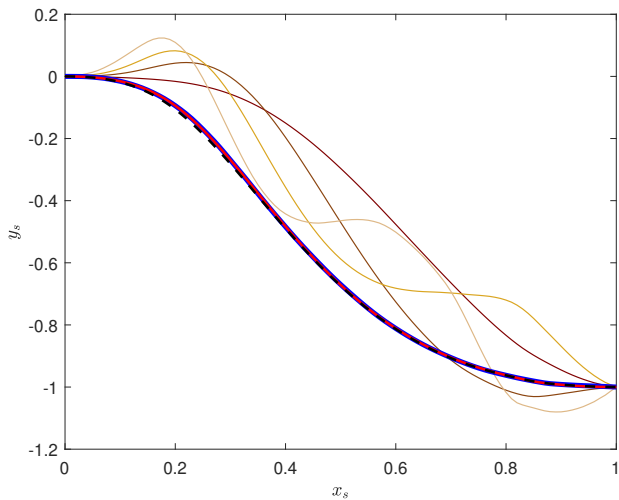}
	\caption{Example of the PCA decomposition for a signal in the
          $x_s-y_s$ space on $s_1$. The four PCA components used to
          reconstruct this segment are shown in shades of brown: the
          first component is the darkest and the fourth component is
          the lightest. The sum of the first four PCA eigenfunctions
          each with the corresponding coefficient, is shown in red,
          while the original signal is in blue (the reconstruction is
          so good that the lines overlap). The signal reconstructed by
          NNs using the predicted PCA coefficients is shown as dashed
          black.   The first four PCA components are enough to represent the signal in each segment.}
		\label{fig:PCA}
\end{figure}

In principle, for a perfect reconstruction of the signal via PCA
decomposition the number of coefficients should be the same as the
size of the database (i.e., $\sim 30,000$ in our case). However, for
our dataset the first four eigenvalues strongly dominate, allowing us
to truncate the basis and use only the first four eigenfunctions to
represent the signal in each segment. Fig.~\ref{fig:PCA} is an
illustration of the PCA decomposition for a signal over $s_1$. We show
a re-normalized signal (blue) and the first four eigenfunctions of the
basis (shades of brown). The red curve shows the sum of the first four
PCA components (each with its corresponding coefficient), reproducing
the true signal nearly perfectly. We quantify the accuracy of the
reconstruction process along each segment by calculating the
r.m.s.\ of the error defined as
  \begin{equation}
  {\rm RMS} = \sqrt{\rm mean\left[  \left( y_{s,\rm sim}(x_s)-y_{s,\rm pred}(x_s)\right) ^2\right] }\ .
  \label{Eq:rms}
  \end{equation}
The mean r.m.s.\ error across all the reconstruction cases is 0.0020 on $s_1$, 0.0058
and 0.0075 on $s_2^p$ and $s_3^p$, respectively, and 0.0045 for $s_2^n$.

\subsubsection{Training of the NNs}
\label{Sec:NNtrained}
Having created the PCA decomposition for each set of astrophysical
parameters from the training dataset of 27,455 cases, we tabulated the
values of the PCA coefficients along with the key points (both the
frequency and the corresponding brightness temperature of each key
point). Using this library, NNs were trained to retrieve the PCA
coefficients along with the values of the key points given an input
set of the astrophysical parameters.   Architecturally, all the NNs described in this section are identical having one hidden layer of 40 neurons and employing the Levenberg-Marquardt algorithm 
 to minimize the
mean-squared error between the true value provided by the training
dataset and the value predicted by the network.

We found that the accuracy of the prediction is improved if we add to
the modeling combinations of the astrophysical parameters that we
expect to map more directly to the 21-cm signal. We made use of the
fact that we know the simulated cosmology. Assuming the standard
collision-less cold dark matter scenario and hierarchical structure
formation, we can infer the mean collapse fraction at every redshift
\citep[$f_{\rm coll}(z)$, the fraction of mass
that is contained in halos of mass above the minimum cooling threshold,][]{Barkana:2004}. We appended five more
parameters to the original set of the seven astrophysical parameters,  bringing the total number of input parameters of each NN to 12. 
The auxiliary parameters include:  $f_* f_{\rm coll}(20)$ which is proportional to the
intensity of the Ly$\alpha$ radiation before the Ly$\alpha$ coupling;
$f_* f_X f_{\rm coll}(15)$ which scales as the intensity of X-ray
radiation before the heating saturation; and $\zeta f_{\rm coll}(10)$
which is a measure of the ionizing radiation at the onset of
reionization. In addition, we added the fraction of X-ray energy above
1~keV and the fraction of X-ray energy above 2~keV ($f_{\rm XR>1~keV}$
and $f_{\rm XR>2~keV}$, respectively) to characterize the X-ray
SED. Lastly, we applied physical cuts on the predicted signal to
assist the NNs. In particular, an upper limit of $T_{\rm max}^{\rm
  hi}=0$ was imposed because $T_{\rm max}^{\rm hi}$ is expected to
always be negative in the range of scenarios considered here. We also
set a lower limit on the signal at the minimum point, $T_{\rm min}$,
in accordance with Eq.~\ref{eq:min}.

In total, predicting a positive/negative signal requires generating
19/13 parameters: four PCA coefficients for each of the three/two
segments plus four/three key points each having two coordinates
(frequency and brightness temperature), minus one degree of freedom
because the value of the brightness temperature at $\nu_{\rm reion}$
is by definition zero. As part of the optimization process we had to
choose between using one network which would predict all the 19/13
output parameters, 19/13 networks each of which would return a single
parameter, or a few NNs predicting groups of the parameters.  We found
that predicting several parameters with a single network sometimes
decreases the error in the predicted signal. However, it also can
result in outliers, i.e., a few cases with very large error. To
minimize the frequency of outliers while preserving the overall
accuracy, we decided to group correlated parameters within the same
network. For instance, the four PCA components of a given segment are
correlated and were computed with a single NN  that has 12 input parameters (and, thus, 12 input neurons), one hidden layer of 40 neurons and four outputs (the PCA coefficients). The outcome of this
prediction is demonstrated in Fig.~\ref{fig:PCA} (dashed black
line). Using Eq.~\ref{Eq:rms} we assessed the performance of the
prediction process and found an r.m.s.\ error of 0.023 on $s_1$,
0.055 and 0.136 on $s_2^p$ and $s_3^p$ respectively, and 0.031 on
$s_2^n$.

Two NNs were trained to predict the coordinates of the critical
points in the $\nu-T_{21}$ space for positive/negative cases.  These NNs have 12 input parameters and 7/5 outputs (4/3  temperature values and 
3/2 frequency coordinates). 
 The accuracy of the reconstruction
of these coordinates is summarized in Table ~\ref{table:succ}. We
found that the algorithm is well tuned to predict the signal from
Cosmic Dawn, with 100\% of cases having better than 5\% accuracy in
the amplitude and the location of the high-redshift maximum, and more
than 98\% of cases having better than 5\% accuracy in the prediction
of the depth and location of the absorption trough. The low-redshift
maximum point is the hardest to predict since it is affected by all
the astrophysical parameters and also the amplitude of the signal at
this point is quite small. In addition, because this feature does not
exist for negative cases, the training dataset which could be used for
($\nu_{\rm max}^{\rm lo}$, $T_{\rm max}^{\rm lo}$) was smaller. The
maximal absolute error obtained when predicting $T_{\rm max}^{\rm lo}$ was
2.3~mK, with 59.87\% of cases returning relative errors smaller than
5\%. In 98.57\% of cases $\nu_{\rm max}^{\rm lo}$ was found to better
than a 5\% error. Finally, the success rate for the prediction of the
timing of reionization was close to 100\%.

 Separate, but architecturally identical (with 12 input parameters, one hidden layer of 40 neurons, one output), NNs were trained to 
predict frequencies at which the neutral fraction is 0.16\% ($\nu_{16\%}$, which we used in Section \ref{Sec:eor}) and 11\% along with  values of the neutral fraction at $z=5.9$, $z=7.08$ and $z=7.54$.

\begin{table*}
\begin{tabular}{|c|c|c|c|c|c|c|c|}
	\hline Key point & $\nu_{\rm max}^{\rm hi}$ & $T_{\rm max}^{\rm hi}$ & $\nu_{\rm min}$ & $T_{\rm min}$ &  $\nu_{\rm max}^{\rm lo}$ &  $T_{\rm max}^{\rm lo}$  & $\nu_{\rm reion}$ \\ 
	\hline Prediction error below 2\% [\%] & 97.44 & 99.59 & 77.02 & 78.88 & 66.88 &    38.54 & 96.68 \\ 
	\hline Prediction error below 5\% [\%] & 100 & 100 & 99.42 & 98.28 & 98.57 & 59.87 & 99.88 \\ 
	\hline 
\end{tabular} 
		\caption{Accuracy in prediction of the key points
                  $\nu_{\rm max}^{\rm hi}$, $T_{\rm
                    max}^{\rm hi}$, $\nu_{\rm min}$,  $T_{\rm min}$, $\nu_{\rm max}^{\rm lo}$, $T_{\rm
                    max}^{\rm lo}$ and       $\nu_{\rm reion}$. The percentage of cases with a
                  relative error below 2\% (5\%) in the prediction is
                  shown in the second (third) row. }
		\label{table:succ}
\end{table*}

\begin{figure*}
	\centering
	\includegraphics[width=3.1in]{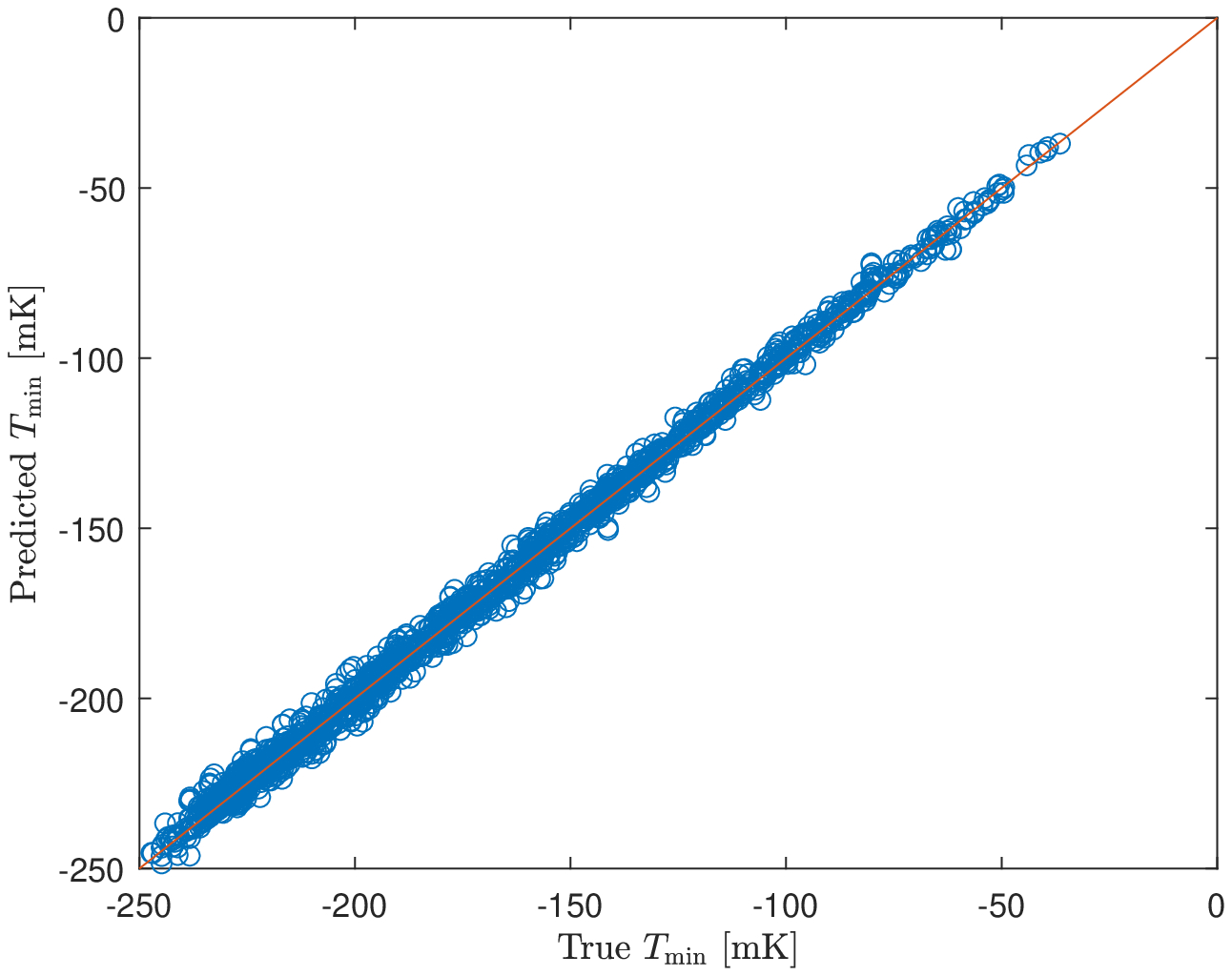}
        	\hspace{0.2in}
	\includegraphics[width=3.5in]{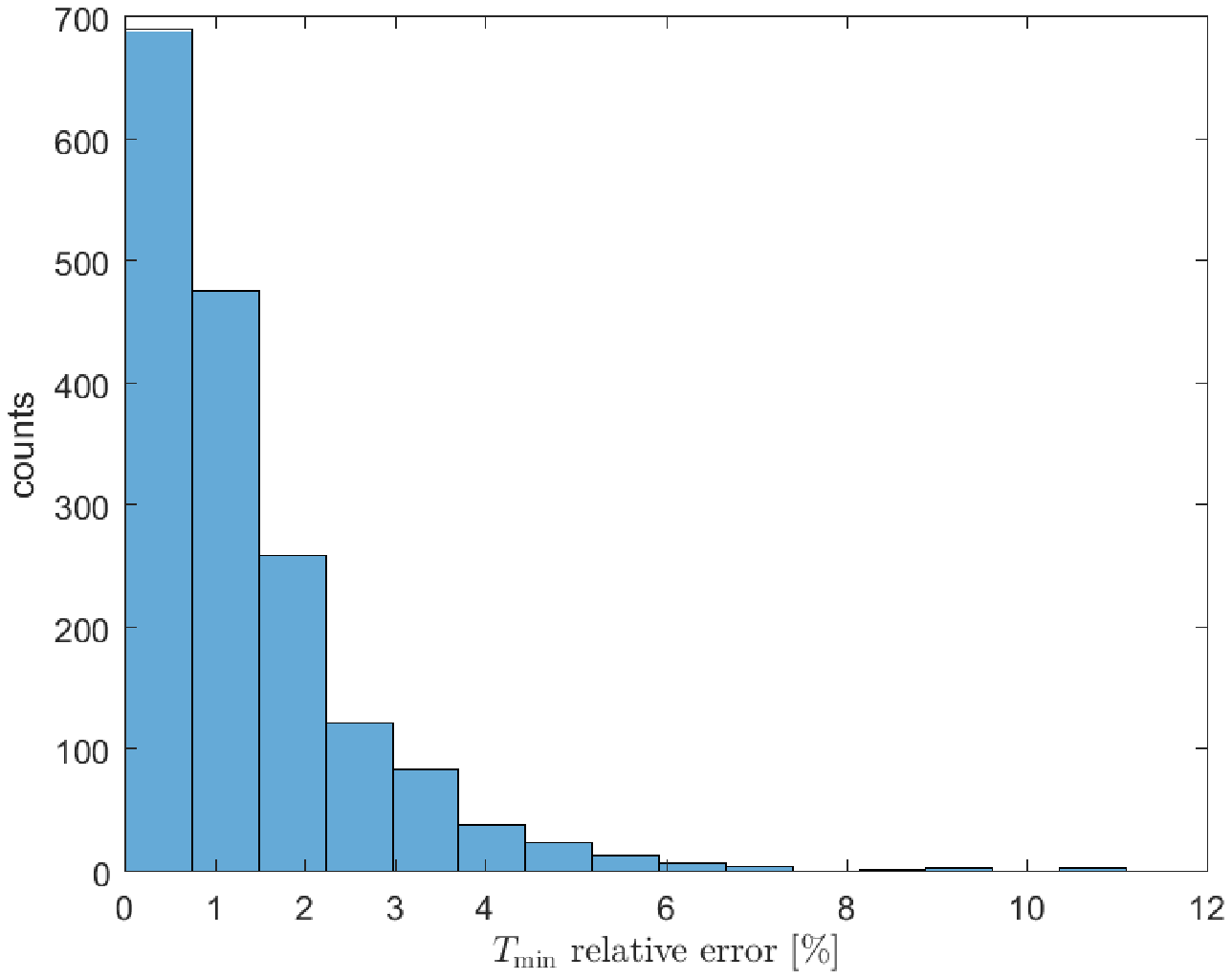}
	\caption{Left panel: Predicted versus true value of the
          amplitude of the absorption feature, shown for 1,743 cases.
          Also shown is the perfect prediction ($Y=X$, solid red
          line). Right panel: Histogram of the relative error in the
          predicted amplitude of the absorption trough.  98.28\% of cases have a relative error of less
than 5\%.}
	\label{fig:Tmin}
\end{figure*}

As an illustration, in Fig.~\ref{fig:Tmin} we show the accuracy of
the algorithm in reconstructing the amplitude of the absorption
feature for 1,743 cases (all our test cases that were not excluded by
the observational constraints in section~\ref{Sec:eor}). The line
$Y=X$ corresponds to a perfect prediction. The scatter shows the error
in this prediction, which is also quantified in the histogram (right
panel). We find that 98.28\% of cases have a relative error of less
than 5\%, while 78.88\% of cases have an error less than 2\% (as
indicated in Table \ref{table:succ}).

\subsection{Prediction Pipeline}
\label{Sec:pipe}
\begin{figure*}
	\centering
	\includegraphics{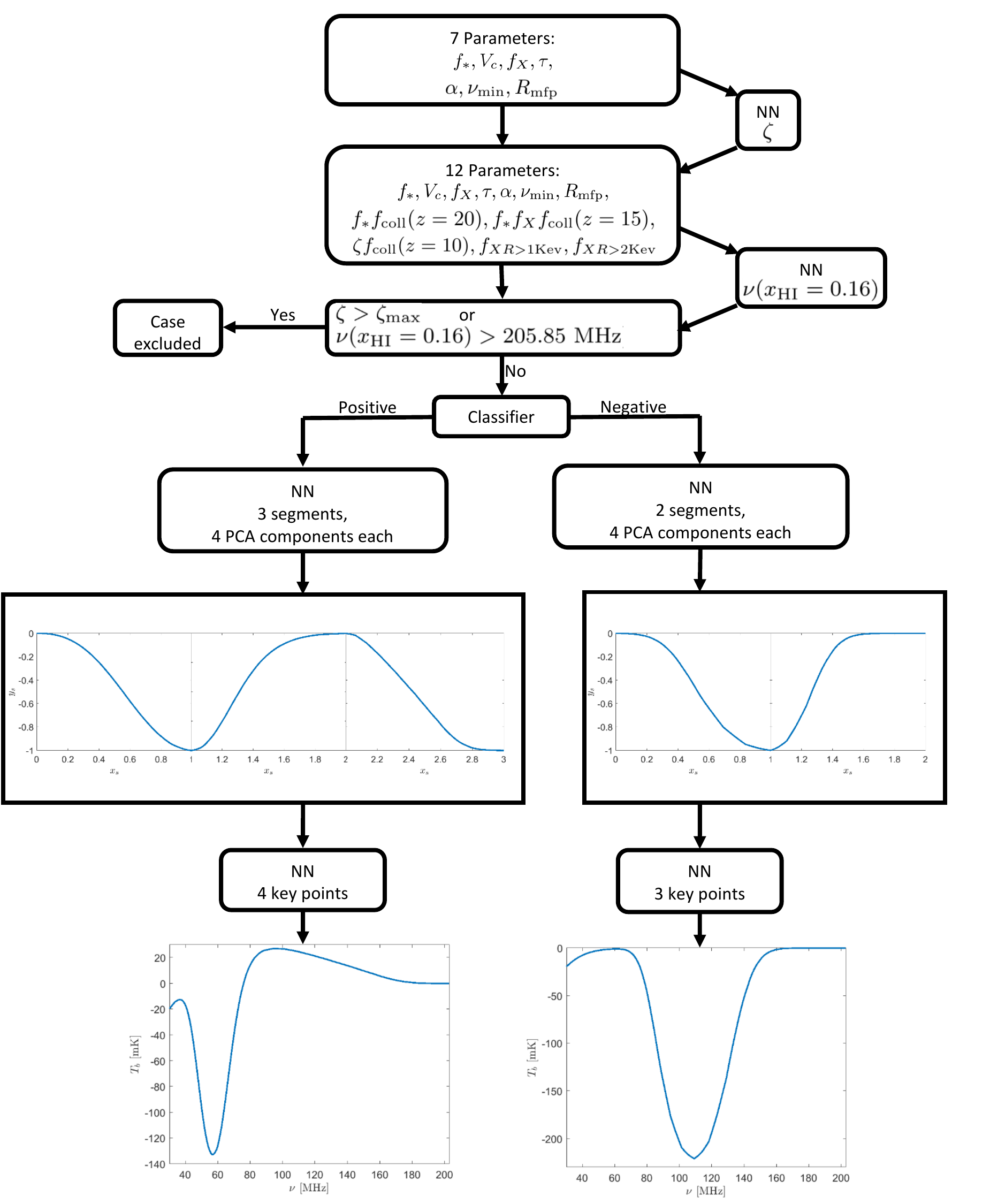}
	\caption{Flowchart of the prediction process as outlined in
          Sec.~\ref{Sec:pipe}.}
	\label{fig:flow}
\end{figure*}

Using the trained NNs the global 21-cm signal is predicted given a set
of 7 input parameters. The complete prediction algorithm is summarized
in Fig.~\ref{fig:flow} and contains the following steps:
\begin{enumerate}
\item Given the seven input astrophysical parameters [$f_*$, $V_c$, $f_X$,
  $\tau$, $\alpha$, $\nu_{\rm min}$, $R_{\rm mfp}$]   trained
  NN (Sec.~\ref{Sec:eor1}) is used to infer the value of the ionizing
  efficiency $\zeta$.
 The algorithm  calculates five auxiliary parameters $f_* f_{\rm
      coll}(20)$, $f_* f_X f_{\rm coll}(15)$, $\zeta f_{\rm
      coll}(10)$, $f_{\rm XR>1~keV}$, $f_{\rm XR>2~keV}$]. 
\item Using the full set of parameters and a NN that predicts $\nu_{16\%}$ the algorithm verifies whether
  this case has a valid EoR history or not (as described in
  Sec.~\ref{Sec:eor}).
\item If the case is valid, the algorithm uses decision trees to
  classify the case and determines if it is expected to have an
  emission feature or not (i.e., whether the case is negative or
  positive, Sec.~\ref{Sec:Class}).
\item Based on the input parameters, NNs (Sec.~\ref{Sec:NNtrained})
  predict the PCA coefficients for each of the segments as well as
  the coordinates of the key astrophysical points (as explained in
  Sec.~\ref{Sec:PCA}).
\item A coordinate transformation (inverse of Eq.~\ref{eq:xs}) is
  performed to return the signal in physical units of mK as a function
  of frequency in MHz.
\end{enumerate}

\subsection{Performance analysis}
\label{Sec:Acc}

In this section we test the overall performance of the emulator,
assessing its accuracy in predicting the global signal for each 
of the test cases in the set of 1,743 signals.

\begin{figure}
	\centering
	\includegraphics[width=3.2in]{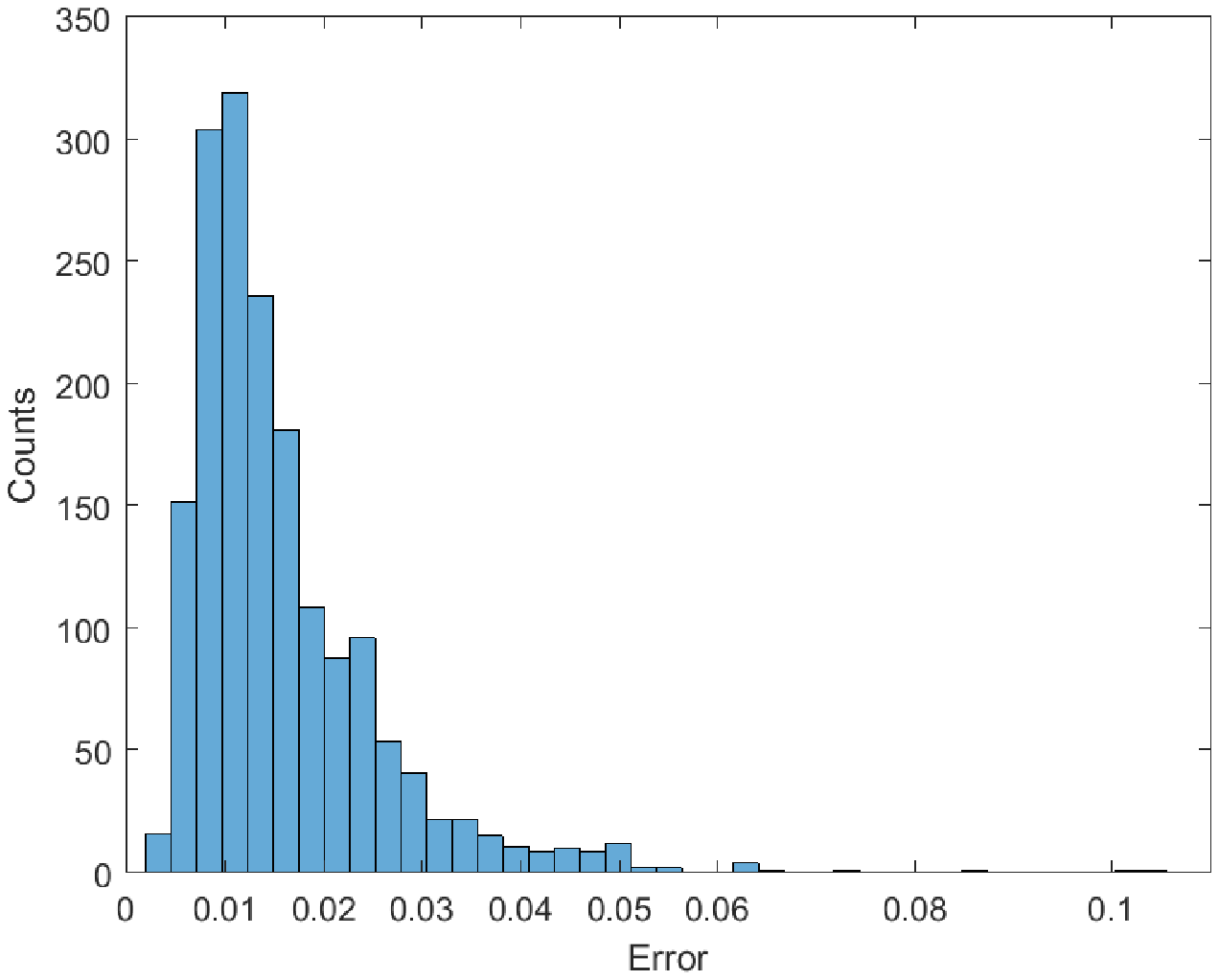}
	\caption{Histogram of the errors in the predicted global 21-cm signals, normalized by their maximum amplitude, as defined in
          Eq.~\ref{eq:error}. The results are shown for the entire test set of 1,743 cases.  The error is
lower than 0.05 for 98.9\% of cases.}
	\label{fig:Error-hist}
\end{figure}

We define the error in the predicted signal, $T_{\rm pred}(\nu)$,
compared to the signal generated by the full simulation for the same
parameter set, $ T_{\rm sim}(\nu)$, as the r.m.s.\ value of the
difference between the two signals, normalized by the maximal
amplitude of the true signal:
\begin{equation}
\label{eq:error}
\text{Error}=\frac{\sqrt{\rm mean\left[  \left( T_{\rm sim}(\nu)-T_{\rm pred}(\nu)\right) ^2\right]  } }{ \max\lvert T_{\rm sim}(\nu)\rvert}.
\end{equation}
Over the entire test set the mean value for the error is 0.0159 and
the median is 0.0130. The histogram of the errors for all the tested
cases is shown in Fig.~\ref{fig:Error-hist}. We find that the error is
lower than 0.05 for 98.9\% of cases.

\begin{figure*}
	\centering
	\includegraphics[width=6in]{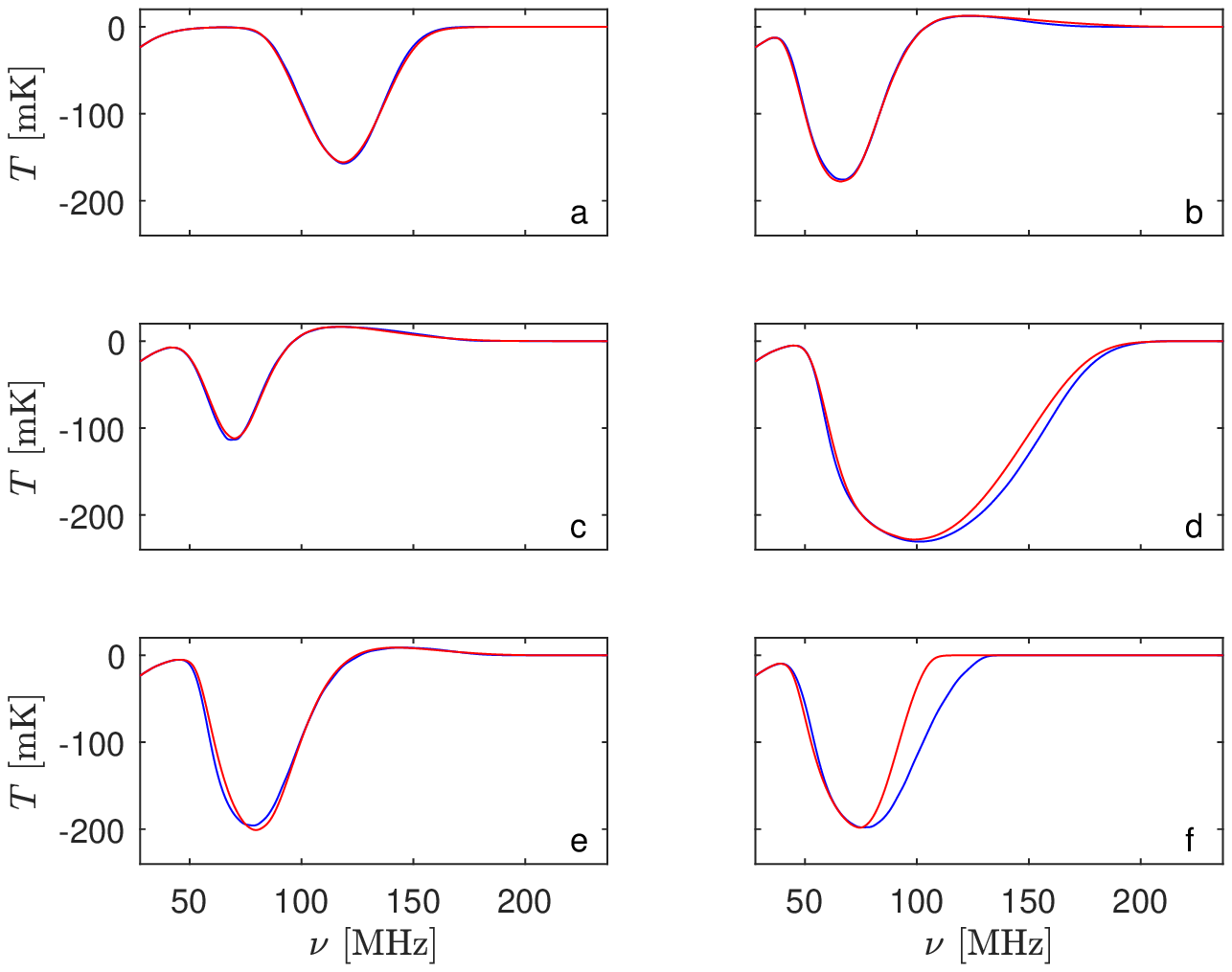}
	\caption{Comparison between the simulated signal (blue) and
          the predicted signal (red). The error and parameters
          [$f_*$,$V_c$,$f_X$,$\tau$,$\alpha$,$\nu_{\rm min}$,$R_{\rm
              mfp}$] of each panel are: (a) 10'th percentile error
          (error = 0.0072; [0.050,76.5,0.001,0.0781,1.3,1,20]); (b)
          median error (error = 0.0130;
          [0.5,4.2,0.1,0.0779,1,0.5,20]); (c) mean error (error =
          0.0159; [0.0285,5.41,3.01,0.0738,1,0.5,34]); (d) 90'th
          percentile error (error = 0.0271;
          [0.2573,21.50,0.0001,0.0701,1.5,0.7,14]); (e) 95'th
          percentile error (error = 0.0349;
          [0.5,24.2,0.02,0.0666,1.5,0.1,35]); (f) the largest error
          (error = 0.1055; [0.3835,7.79,0.0045,0.1,1.5,0.2,48.30]).  }
	\label{fig:ex}
\end{figure*}

To illustrate the performance of the emulator we show several specific
cases in Fig.~\ref{fig:ex}: (a) the case with a 10'th percentile error
(i.e., 10\% of the cases have smaller error), with an error of 0.0072,
(b) the median error of 0.013, (c) the mean error of 0.0159, (d) 90'th
percentile error of 0.0271, (e) 95'th percentile error of 0.0349, and
(f) the largest error of 0.1055. Visually, the cases with the mean and
median errors (top right and middle left panels) are in excellent
agreement with the simulated signal.

\subsection{Limitations}

 {\sc 21cmGEM} is designed to cover a wide range of redshifts ($5-50$) and was optimized to return a small mean error over the entire range. This is both an advantage and a disadvantage. The reionization parameters $\tau$ and $R_{\rm mfp}$ only affect the low-redshift portion of the signal (at $z\lesssim 10$) where the amplitude of the signal is very low (compared to the deep absorption trough at Cosmic Dawn). Therefore, if {\sc 21cmGEM} were used as a part of MCMC to recover these parameters from data, large errors would be expected.
 For instance, $R_{\rm mfp}=70$ Mpc results in a slightly  faster end to reionization, compared to 50 Mpc. However, the difference between the global signals with $R_{\rm mfp}=70$ Mpc and 50 Mpc is very small. E.g., for a model  with  $V_c=16.5$ km s$^{-1}$, $f_* = 0.05$, $f_X = 1$, hard SED, and $\tau  = 0.073$, the error  is $ 9.7\times 10^{-4}$, which is  much smaller than the typical precision of {\sc 21cmGEM} with the median value of r.m.s. of 0.01, as discussed in Section \ref{Sec:Acc} of this paper. A related shortcoming is the precision of    {\sc 21cmGEM} in reconstructing the emission feature and the large error on $T_{\rm max}^{\rm lo}$ (as we discussed in Section \ref{Sec:NNtrained}).

\section{Summary and Conclusions}
\label{Sec:Summ}
In this paper we have presented a database of 29,641 global 21-cm
signals generated over the widest possible space of seven
astrophysical parameters that include the star formation efficiency,
minimum cooling mass, X-ray radiation efficiency, the slope and the
low energy cut-off of the X-ray spectrum, the mean free path of the
ionizing photons, and the CMB optical depth. The parameter space is
constrained by the observations of the CMB and quasar absorption lines
as well as by the maximum possible ionizing efficiency (corresponding
to massive metal-free stars). 

We used this dataset to verify the
consistency relations between the astrophysical parameters and the
properties of the global 21-cm signal first reported in our previous
paper \citep{Cohen:2017}, finding a good agreement in all relations
except for the value of $J_\alpha/\epsilon$ at $z_{\rm min}$ which
shows much larger scatter due to the wider selection of X-ray spectra
considered here. In particular, there remains a tight predicted
relationship between the brightness temperature and the observed
frequency of the high-redshift maximum point (Fig.~\ref{fig:hmax}); a
measurement of this point can be used to infer the Ly$\alpha$
intensity at that time, though with significant scatter
(Fig.~\ref{fig:hmaxJA}). Also, the brightness temperature and observed
frequency of the low-redshift maximum point follow a tight relation,
which can be used to estimate the X-ray intensity
(Fig.~\ref{fig:lmax}).

We utilized the database to develop and test {\sc 21cmGEM} which,
given a set of astrophysical parameters, predicts the global 21-cm
signal.  Additional outputs include values of the neutral fraction at $z=5.9$, $z=7.08$ and $z=7.54$ along with frequencies at which the neutral fraction is 0.16\% and 11\%. The crucial elements of the emulator are: 
\begin{enumerate}
\item Smoothness of the output signal is guaranteed by construction.
\item The database can be divided into two categories: signals that
  have an emission feature and signals that are only seen in
  absorption. The classification is done using bagged decision trees.
\item We train neural networks to predict the cosmological signal
  based on the seven astrophysical parameters.  Each signal is broken
  into a few segments separated by the key astrophysical points and is
  decomposed into a basis of smooth orthogonal functions using
  PCA. Because PCA ranks the eigenfunctions by variance in 
  decreasing order, most of the information is encoded in the first
  few terms.  This allows us to reduce the dimensionality and use only
  the first four functions of the basis. Neural networks are used to
  predict the PCA coefficients as well as the two ends of each segment
  given a set of astrophysical parameters.
\item The algorithm also checks whether the case satisfies current
  constraints on reionization. The constraints that are taken into
  account include limits on the total CMB optical depth, the upper
  limit on the ionization efficiency of stars, and the upper limit on
  the neutral fraction at $z\sim5.9$ derived from the absorption
  profile of high-redshift quasars. Using these conditions the
  minimum circular velocity of starforming halos as well as the star
  formation efficiency can be constrained. We find a lower limit of
  $V_c\sim26$ km s$^{-1}$ ($\sim4\times M^{\rm atomic}_{\rm min}$) for
  an optical depth of 0.055.
\end{enumerate}
 The algorithm was trained on 27,455 simulated signals, and an
 additional 2,186 cases were used as the test set. The predicted signal
 has an r.m.s.\ error of 0.0159, corresponding to 1.59\% of the signal amplitude, with 98.9\% of cases having errors
 lower than 0.05. The algorithm is efficient, with running time per
 parameter set of 0.16~sec (while one full simulation run typically takes a few hours on a single core). This tool can be used in the fitting
 process (e.g., MCMC) to constrain the high-redshift parameter space
 using the data of global signal experiments. We have used it recently
 with the data from EDGES High-Band  \citep{Monsalve:2019}. {\sc 21cmGEM} and the training and testing datasets are available online at  \url{https://www.ast.cam.ac.uk/~afialkov/Publications.html}.

\section{Acknowledgments}
We acknowledge the usage of the Harvard Odyssey cluster. Part of this
project was supported by the Royal Society University Research
Fellowship of AF. RB and AC acknowledge Israel Science Foundation
grant 823/09 and the Ministry of Science and Technology, Israel. This
project/publication was made possible for RB through the support of a
grant from the John Templeton Foundation. The opinions expressed in
this publication are those of the authors and do not necessarily
reflect the views of the John Templeton Foundation. RB was also
supported by the ISF-NSFC joint research program (grant
No. 2580/17). RAM was supported by the NASA Solar System Exploration
Virtual Institute cooperative agreement 80ARC017M0006.

\end{document}